\documentstyle[emulateapj,colordvi]{article}
\newcommand{\kms}{km~s$^{-1}$}
\newcommand{\lya}{Ly$\alpha$}
\def\o6{\ion{O}{6}}

\input epsf

\begin{document}
\title{THE HOT INTERGALACTIC MEDIUM -- GALAXY CONNECTION: TWO STRONG
O~VI ABSORBERS IN THE SIGHTLINE TOWARD PG~1211+143}
\author{
JASON TUMLINSON\altaffilmark{1},
J. MICHAEL SHULL\altaffilmark{2,3},
MARK L. GIROUX\altaffilmark{4}, \&
JOHN T. STOCKE\altaffilmark{2}}

\altaffiltext{1}{Department of Astronomy and Astrophysics,
                 University of Chicago,
                 5640 S. Ellis Ave., Chicago, IL 60637}
\altaffiltext{2}{Department of Astrophysical and Planetary Sciences
                and CASA,
                 University of Colorado, Boulder, CO 80309}
\altaffiltext{3}{Also at JILA, University of Colorado and
                 National Institute for Standards and Technology}
\altaffiltext{4}{Department of Physics, Astronomy, and Geology, Box 70652,
                 East Tennessee State University, Johnson City, TN 37614}

\begin{abstract}
We present {\em HST}/STIS and {\em FUSE} spectra of the QSO
PG~1211+143 ($z_{em} = 0.081$) and a galaxy survey of the
surrounding field. This sightline shows two strong intergalactic
absorption systems at $cz \simeq 15,300$ and 19,300 \kms.  This
sightline addresses the nature and origin of the \ion{O}{6}
absorbers, and their connection to galaxies.  We explore the
relationship of these absorbers to the nearby galaxies and compare
them to other \ion{O}{6}-bearing absorbers in diverse
environments. At 15,300 \kms, we find four distinct \ion{H}{1}
components and associated \ion{C}{2}, \ion{C}{3}, \ion{C}{4},
\ion{Si}{2}, \ion{Si}{3}, \ion{Si}{4}, \ion{N}{5}, and \o6, lying
near a spiral-dominated galaxy group with a bright member galaxy
137$h_{70}^{-1}$ kpc from the sightline.  The observed ions of C,
Si, and N are likely to be photoionized, but the \ion{O}{6} is
more consistent with collisional ionization. The ion ratios in
this absorber resemble the highly-ionized Galactic HVCs; it may
also trace the hot intragroup medium gas or the unbound wind of an
undiscovered dwarf galaxy.  At 19,300 \kms, we find five
\ion{H}{1} components and associated \ion{C}{3}, \ion{Si}{3}, and
collisionally-ionized \o6 lying 146$h_{70}^{-1}$ kpc from an
isolated galaxy.  The properties of the \ion{O}{6}-bearing gas are
consistent with an origin in strong shocks between low-metallicity
gas ($\geq 2 - 6$ \% solar) and one or more of the warm
photoionized components.  It is likely that these absorbers are
related to the nearby galaxies, perhaps by outflows or gas
stripped from unseen satellite galaxies by interactions. However,
we cannot reject completely the hypothesis that they reside in the
same large-scale structure in which the galaxies are embedded but
are otherwise not directly related.
\end{abstract}


\section{INTRODUCTION}

The uncertain role of hot ($10^{5-7}$ K) gas in the intergalactic
medium (IGM) and galaxy halos motivates us to seek out
highly-ionized tracers in different environments. The most
accessible tracer in this temperature range is the \o6\
$\lambda\lambda$1032,1038 doublet. With ionization energy of 114 eV
required for its production, \o6\ arises in gas at $T = 10^{5 - 6}$
K in collisional ionization equilibrium (CIE; Sutherland \& Dopita
1993). Two important scientific goals motivate the current
observational searches for \ion{O}{6} absorption in the Galactic
halo, Local Group, and IGM. First, \o6\ may help to locate some of
the predicted ``warm-hot intergalactic medium'' (WHIM) thought to
hold the 30 -- 40\% of baryons not yet accounted for at low redshift
(Fukugita, Hogan, \& Peebles 1998; Stocke, Shull, \& Penton 2004b).
Theory predicts that a substantial fraction of the baryons at low
$z$ are shock-heated to $T > 10^5$ K by gravitational infall into
galaxies, groups, and other large-scale structures (Cen \& Ostriker
1999; Dav\'{e} et al. 1999).  \ion{O}{6} could reveal the existence
and location of some of these ``missing baryons''. Second, the large
{\em FUSE} survey of \o6\ (Sembach et al. 2003) suggests that hot
gas forms an important component of the Milky Way halo and/or Local
Group.  Searches for \ion{O}{6} near other galaxies could generalize
this result.

Hydrodynamical simulations (Cen \& Ostriker 1999; Dav\'{e} et al.
1999) of large-scale structure predict that the low-redshift IGM is
approximately equally divided by mass into three distinct phases:
gas that has condensed to form stars and galaxies, gas that remains
quiescently photoionized in the diffuse IGM ($T \sim 10^{4-5}$~K),
and hotter gas ($T \sim 10^{5-7}$~K) that has fallen into the denser
regions of the IGM where it is shock-heated and ionized.  The cold,
collapsed phase is well-understood from decades of studying
galaxies. At low redshift, the photoionized phase has been studied
extensively by {\em HST} (Bahcall et al. 1993; Penton, Shull, \&
Stocke 2000) and found to contain 29 $\pm$ 4\% of the baryons in the
local universe (Penton, Stocke, \& Shull 2004, and references
therein). The hot phase can be traced by the resonance lines of the
highly-ionized species \o6, \ion{O}{7}, and \ion{O}{8}. Tripp,
Savage, \& Jenkins (2000), Savage et al. (2002), and Danforth \&
Shull (2004) have found \o6\ absorbers in numbers that suggest that
5 - 10\% of the baryons reside in the $10^{5-6}$~K phase, but X-ray
searches for \ion{O}{7} and \ion{O}{8} are just beginning with {\em
Chandra} (Fang et al. 2002; Nicastro et al. 2002; Nicastro et al.
2004) and do not yet provide accurate baryon densities.

Another promising reservoir of hot gas lies in the extended halos of
galaxies, as suggested by the discovery of widespread high-velocity
\ion{O}{6} absorption in the environs of the Milky Way and/or Local
Group by {\em FUSE} (Sembach et al. 2003).  This survey found 84
distinct high-velocity \ion{O}{6} systems in 102 surveyed extragalactic
sightlines.  While some of these highly-ionized clouds trace previously
known \ion{H}{1} structures, most do not, and their column densities and
ionization conditions are too diverse to have a common origin. Instead,
Sembach et al. (2003) and Collins et al. (2004) proposed a combination
of photoionized and collisionally ionized models to explain the
\ion{O}{6}. For the \ion{O}{6} without \ion{H}{1} 21 cm emission, they
favor a model in which \ion{O}{6} is collisionally ionized in interfaces
between a hot ($\sim 10^7$ K), extended ($\gtrsim 70$ kpc, the distance
to the Magellanic Stream) Galactic halo and cooler infalling material
(Fox et al. 2004).  In this model, the HVC \ion{O}{6} is an indirect
indicator of the hotter gas rather than a major halo component in its
own right.  This key result also indicates that regions within 100 kpc
of galaxies are fruitful places to search for the signatures of hot
baryons participating in the formation and interaction of galaxies.

An alternative location for hot baryons is in the potential wells of
galaxy groups, as proposed by Mulchaey et al. (1996). Intragroup gas
in small, spiral-dominated groups is expected to have the right
metallicity and temperature to show \ion{O}{6} as a direct signature
of hot baryons. Although this prediction has yet to be confirmed,
Shull, Tumlinson, \& Giroux (2003) found \ion{O}{6} associated with
a galaxy group on the sightline to PKS~2155-304. They suggested that
this \ion{O}{6} is collisionally ionized in the interface between
infalling cool gas and the hot intragroup medium, such that
\ion{O}{6} serves as an indicator of the hotter WHIM gas, as it
appears to in the Galactic halo.

Although the expected signatures of hot, shock-heated IGM gas have
been identified in QSO absorption lines, their connection to
galaxies is still largely unknown. Correlations between galaxies and
Ly$\alpha$ clouds have been described in a statistical sense
(Lanzetta et al. 1995; Chen et al. 2001; Bowen, Pettini, \& Blades
2002; Penton, Shull, \& Stocke 2000; Penton, Stocke, \& Shull 2002,
2004). The locations of Ly$\alpha$ clouds (the warm phase) and the
WHIM relative to galaxies can be examined at a phenomenological
level with numerical simulations (Dav\'{e} et al. 1999). However,
observational evidence linking Ly$\alpha$ clouds and the WHIM to
galaxies or large-scale structure is rare, owing to two technical
barriers to large samples. First, obtaining high-resolution QSO
spectra that cover all interesting IGM absorption lines is
difficult. Second, these studies often must observe only the
UV-brightest targets, which are usually selected without regard to
the foreground galaxy populations. This factor entails intensive
galaxy field surveys to follow up the spectroscopic studies. Both
the high-quality spectra and complete galaxy redshift surveys are
necessary to draw causal connections between IGM absorbers and
galaxies. Once obtained, the data often yield inconclusive results
on the origin of the absorbing material owing to line blending or
insufficient column-density sensitivity. Thus, even though some IGM
material shows evidence of influence from galaxies (generally metal
enrichment), it is difficult to relate absorbers to individual
galaxies or the large-scale structure in any general way. While most
low column density Ly$\alpha$ absorbers are plausibly related to
large-scale structure (Penton et al. 2002; Rosenberg et al. 2003), a
compelling case for individual cloud/galaxy associations can be made
for only a few of the strongest Ly$\alpha$ absorbers (for example,
3C~273, Stocke et al. 2004a and PKS 2155-304, Shull, Tumlinson, \&
Giroux 2003).

In this study, we examine two strong, \ion{O}{6}-bearing IGM
absorption-line systems toward PG~1211+143 and analyze their
connections to galaxies. With $W_\lambda = 1130$ m\AA\ at 15,300
\kms\ and $880$ m\AA\ at 19,300 \kms, these systems are among the
strongest Ly$\alpha$ absorbers at low redshift (Penton, Shull, \&
Stocke 2004). These systems address two major issues of recent
concern to studies of the IGM. First, what is the nature and origin
of the \ion{O}{6} absorbers? Second, what connection do these
absorbers have to galaxies?

We describe our data and analysis techniques in \S~\ref{datasection}. In
\S~\ref{modelsection} we discuss photoionization and
collisional ionization models of the absorbers.  In
\S~\ref{originsection} we compare these systems with other well-studied
IGM and Galactic halo absorbers and discuss physical models for their
origins.  In \S~\ref{conclusionsection} we summarize our specific results
on these absorbers and draw general conclusions about the connections
between galaxies and the \o6-bearing IGM.

\section{DATA AND ANALYSIS} \label{datasection}

\subsection{STIS Data}

The QSO PG~1211+143 ($V = 14.63$, $z_{em} = 0.081$) was observed for
67 ksec on 2002 February 4 -- 8 with the {\em Hubble Space
Telescope}/Space Telescope Imaging Spectrograph (E140M grating, $R =
44,000$; Program ID GO8571). The data were obtained from the STScI
archive and calibrated by the standard {\sc CALSTIS} calibration
pipeline. The 16 sub-exposures were co-added with weighting
proportional to their individual signal-to-noise ratios.  The data
span wavelengths from 1146 \AA\ to 1726 \AA\ in 44 echelle orders
with a signal-to-noise ratio of 10 to 25 per 7 \kms\ E140M
resolution element.

\begin{figure}
\centerline{\epsfxsize=1.0\hsize{\epsfbox{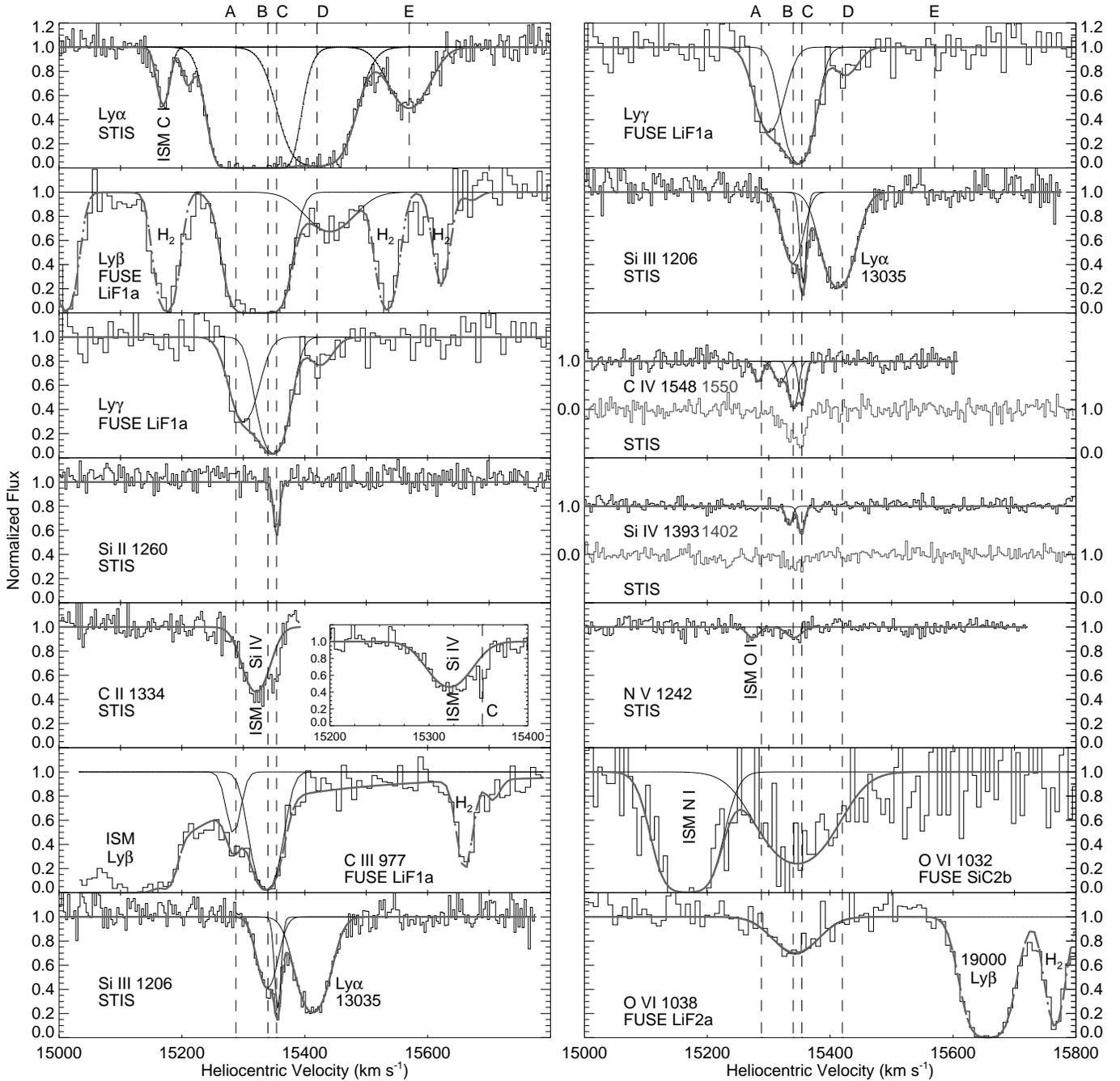}}} \caption{The
absorption lines in the 15,300 \kms\ system, aligned in
heliocentric velocity space. The panels are marked with the
species of interest and the source of the data ({\em FUSE} or
STIS). The components from Table~\ref{15table} are marked in
vertical dashed lines, and labeled at the top. Black profiles are
the individual model components, and thick gray lines trace the
overall model line profiles. Contaminating Galactic absorption is
labeled (H$_2$, Ly$\beta$, C~I, N~I, O~I, Si~IV, including model
fits where necessary). The panel with C~II $\lambda$1334 shows the
same line on two adjacent STIS echelle orders, where Component C
in C~II $\lambda$1334 lies in the wing of Galactic Si~IV 1402. In
both cases, the plotted model was derived from a fit to the
Galactic Si~IV $\lambda$1393 line and then applied directly to the
$\lambda$1402 profile to show the excess absorption by C~II at
15,353 \kms. In the right column, H~I Ly$\gamma$ and Si~III
$\lambda$1206 are repeated to ease visual comparisons of the
high-ion components with the lower ions and \ion{H}{1}. The N~V
$\lambda$1238 line is not shown because it lies directly
underneath the strong Galactic ISM O~I $\lambda$1302 line. For
C~IV and Si~IV, we plot both lines of the doublet; the stronger
line is plotted above and corresponds to the left side axis, while
the weaker and bottom line corresponds to the right hand axis.
\label{stack15}}
\end{figure}

The calibrated STIS spectra were examined for \ion{H}{1} and metal
absorption lines at the redshifts of the known Ly$\alpha$
complexes at $\simeq$ 15,300 and $\simeq$ 19,300 \kms, which were
discovered with STIS G140M (Penton, Stocke, \& Shull 2004).
Figures~\ref{stack15} and \ref{stack19} display
continuum-normalized absorption-line profiles for the 15,300 \kms\
and 19,300 \kms\ systems, respectively, in heliocentric velocity
space.  In Figure~\ref{stack15} both lines of the \ion{O}{6},
\ion{C}{4}, and \ion{Si}{4} doublets are plotted. For the weaker
lines in the \ion{C}{4} and \ion{Si}{4} doublets, the scale is
drawn on the right axis and the data are shifted downward, such
that the continuum line for the lower spectrum serves as the
zeropoint of the upper spectrum. The component groups discussed in
\S~3 are marked with vertical dashed lines, and model absorption
profiles are overlaid with solid curves. The component fitting and
groupings are explained in \S~2.3 and \S~3.

\subsection{FUSE Data}

{\em FUSE} observed PG~1211+143 for 52.3 ksec on 2000 April 25
(observation P1072001). The {\em FUSE} satellite and its
spectrograph performance are described by Moos et al. (2000),
Sahnow et al. (2000), and in the {\em FUSE} Observer's
Guide\footnote{Available online at
http://fuse.pha.jhu.edu/support/guide/guide.html}. {\em FUSE} has
an effective resolution that varies from R = 15,000 -- 20,000 over
910 - 1187 \AA.  We calibrated the data with {\sc CALFUSE} 2.2.1,
the latest version available at the time of our analysis. The
software applied a screen for detector bursts, extracted the
one-dimensional spectrum from the two-dimensional detector images,
and applied calibrations for wavelength and flux. We did not
depart from the standard {\sc CALFUSE} procedure, and we did not
apply orbital day/night screening. We used all 37 sub-exposures in
observation P1072001. The final effective exposure time was
reduced to 51.7 ksec when detector bursts were excised. The {\em
FUSE} SiC1 channel was not aligned with the other channels during
the observation and provided no data. We applied small velocity
shifts to the calibrated data to correct for the $\lesssim 10$
\kms\ systematic fluctuations in the {\em FUSE} wavelength
solution. These corrections were derived by aligning interstellar
\ion{Fe}{2} and \ion{Si}{2} lines detected in the STIS and {\em
FUSE} data.

\subsection{Line Fitting and Analysis}

\begin{figure}
%\plotone{stack19.eps}
\centerline{\epsfxsize=0.5\hsize{\epsfbox{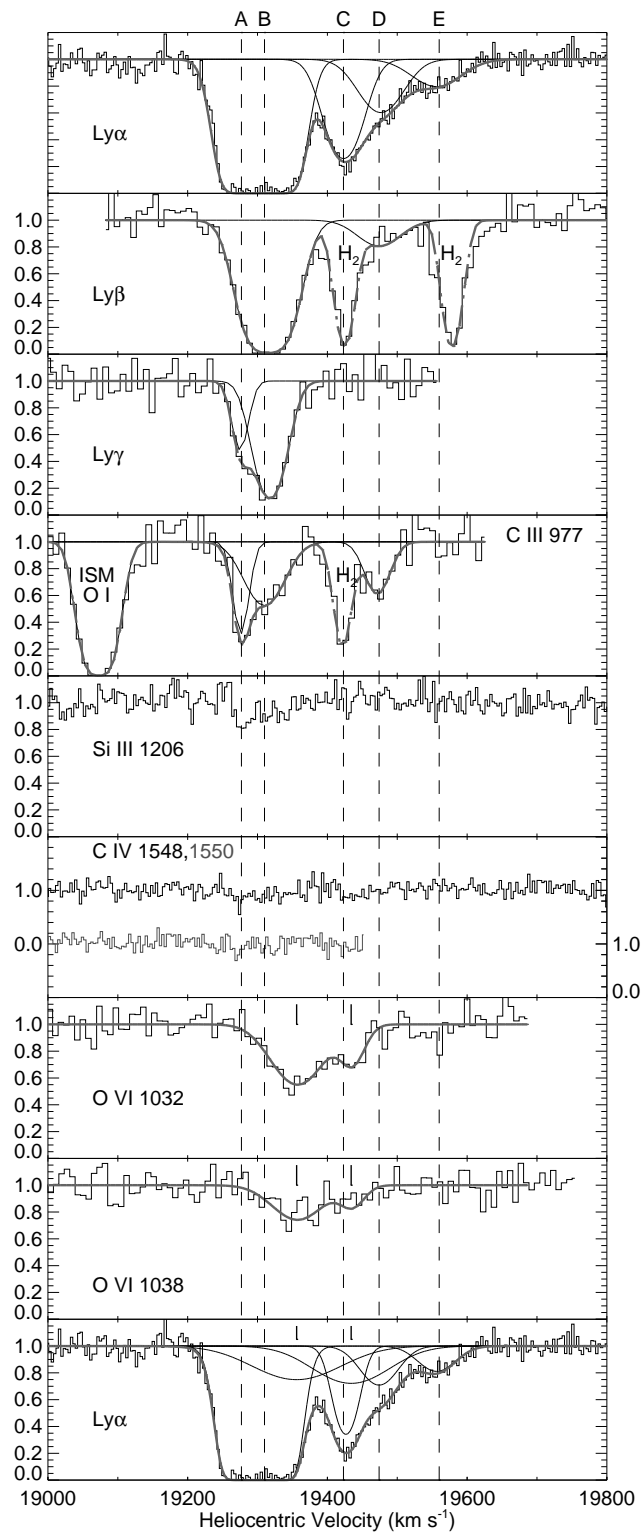}}} \caption{The
absorption lines in the 19,300 \kms\ system, aligned in
heliocentric velocity space. The components from
Table~\ref{19table} are marked in vertical dashed lines, and
labeled at the top. Black profiles are the individual model
components, and thick gray lines trace the combined model
profiles. Contaminating Galactic absorption is labeled where it
exists (O~I, H$_2$). Si~III $\lambda$1206 does not show a model
fit because its column density was obtained by direct integration
of the data. For C~IV we plot data covering both lines of the
undetected doublet; the stronger line is plotted above and
corresponds to the left side axis, while the weaker and bottom
line corresponds to the right hand axis. The bottom panel shows a
model incorporating possible \ion{H}{1} associated with
collisionally ionized O~VI components into the Ly$\alpha$ profile.
The two additional components are included here and the line
centroids of these two components are marked with ticks in the
lower three panels (see \S~3.2). \label{stack19}}
\end{figure}

We analyzed the {\em HST}/STIS and {\em FUSE} absorption-line data
with a combination of profile-fitting and direct-integration
techniques. Absorption lines of interest were identified as lying
within $\pm$ 500 \kms\ of 15,300 and 19,300 \kms. The blended
profiles were fitted as multiple components described individually
by the column density $N$ (cm$^{-2}$), heliocentric velocity $v
\equiv cz$ (\kms), and doppler $b$ parameter (\kms), together with a
continuum composed of Legendre polynomials up to third order. The
lines and continuum were fitted simultaneously by minimizing the
$\chi^2$ statistic and including Galactic interstellar molecular
hydrogen (H$_2$) and atomic contamination where they exist; these
contaminating lines are labeled in Figures~\ref{stack15} and
\ref{stack19}. Individual velocity components in blended profiles
were fitted simultaneously, but the Lyman series lines in each
system were fitted separately to give relatively independent
measures of the parameters. This provides a helpful check on
uncertainties caused by blending, saturation, and the possibility of
unseen components in the stronger lines. In most cases we adopt
$N$(\ion{H}{1}) from the profile fits from the higher Lyman lines
(Ly$\beta$ or Ly$\gamma$) because these are the only profiles where
the components can be confidently separated. In some cases, certain
parameters were fixed to help constrain the fit; these exceptions
are noted below in \S~3. The tabulated error bars correspond to 1
$\sigma$ confidence intervals on the parameters, obtained by finding
the maximum and minimum parameter values giving $\Delta\chi^{2}$ = 1
from the best fit with all other parameters re-optimized.  For both
systems, the fitted \ion{H}{1} columns were checked for consistency
against a curve-of-growth analysis based on Ly$\alpha$ --
Ly$\gamma$.

For {\em FUSE} data, we assume instrumental broadening by a Gaussian
line spread function (LSF) with a full-width at half maximum (FWHM)
corresponding to $R = 15,000$. For STIS data, we use the
non-Gaussian LSF for the E140M grating and the $0.2'' \times 0.2''$
aperture ($R = 44,000$; see the STIS Instrument
Handbook\footnote{Available online at
http://www.stsci.edu/hst/stis/}). Formal doppler $b$ parameters can
be fitted down to the level at which the intrinsic FWHM ($\Delta
v=1.67b$) becomes comparable to the instrumental resolution. For
STIS, we quote formally fitted linewidths down to $\simeq 5$ \kms,
but these narrow linewidths carry the additional uncertainty of
being near the ``effective'' resolution including the broad wings of
the STIS LSF. We can determine $b \gtrsim 12$ \kms\ from {\em FUSE}
data based on our assumption that $R=15,000$ with a Gaussian
line-spread function ({\rm FUSE} can achieve up to $R=20,000$ when
subexposures are carefully aligned, but we assume $R = 15,000$ here
to be conservative). However, there are some lines in the FUSE data
that yielded formal best fits with smaller $b$, and other lines that
were fitted with fixed $b < 12$ \kms\ based on linewidths for other
species determined from the STIS data. These special cases are noted
in the tables and in the discussion of these lines in the text.
These linewidth uncertainties do not generally have adverse effects
on the analysis.

The profile-fitting analysis generated the line parameters discussed
in \S~3.  Consistent fit results across species are grouped according to
velocity and are marked in Figures~\ref{stack15} and \ref{stack19}. These
component groups are assumed to arise in the same gas and are interpreted
as such in the sections below. See \S~3 for details.

\subsection{Galaxy Field}

We compiled the available galaxy redshift data on the PG~1211+143
field to enable detailed comparisons between the IGM absorption-line
systems and nearby galaxies (Table~\ref{galtable}).  The galaxy
redshifts are drawn from the approximately 600,000 galaxies in the
revised Center for Astrophysics redshift survey (Huchra et~al. 1992;
March 2004 version) and the several hundred galaxy redshifts near
bright QSO sightlines obtained by McLin (2002; see also Stocke
et~al. 2005, in preparation).  The combination of these galaxy
catalogs provides a pencil-beam redshift survey that is complete to
$m_B$ = 15.5 for arbitrarily large impact parameter and to $m_B$ =
19 for a 2.5$'$ impact parameter to PG~1211+143. While the former
magnitude limit, based on the CfA redshift survey, is a standard
Zwicky blue-band magnitude, the deeper survey utilized red
magnitudes corrected to $m_B$ by assuming $B - R = 0.7$ (similar to
an Scd galaxy). This conservative color assumption ensures that the
quoted magnitude limit is appropriate for even quite blue galaxies.
The Stocke et~al. (2005, in preparation) magnitudes are 0.65
magnitudes brighter in the mean than measured blue magnitudes for a
large sample of galaxies observed by several surveys.  To be
consistent, we have corrected the Stocke et~al. (2005) magnitudes
using the observed spectral energy distributions for the three
galaxies that are not listed in the CfA catalog.

We do not believe that PG~1211+143 itself affects the quoted
magnitude limit. At $cz$ = 19,300 \kms, an angular separation of
$1''$ corresponds to 1.2 $h_{70}^{-1}$ kpc and our magnitude limit
corresponds to a $\sim 0.3L^*$ galaxy. Therefore, we do not believe
that we would have missed a galaxy at this limit or brighter because
it was obscured by the QSO itself. At an impact parameter of
$\lesssim 1''$, a much higher $N$(\ion{H}{1}) absorber (i.e., a
damped Ly$\alpha$ system) would be expected because the sightline
would intercept the gaseous disk in this case (see e.g., Bowen,
Tripp, \& Jenkins 2001).

\begin{figure*}[t]
\centerline{\epsfxsize=1.0\hsize{\epsfbox{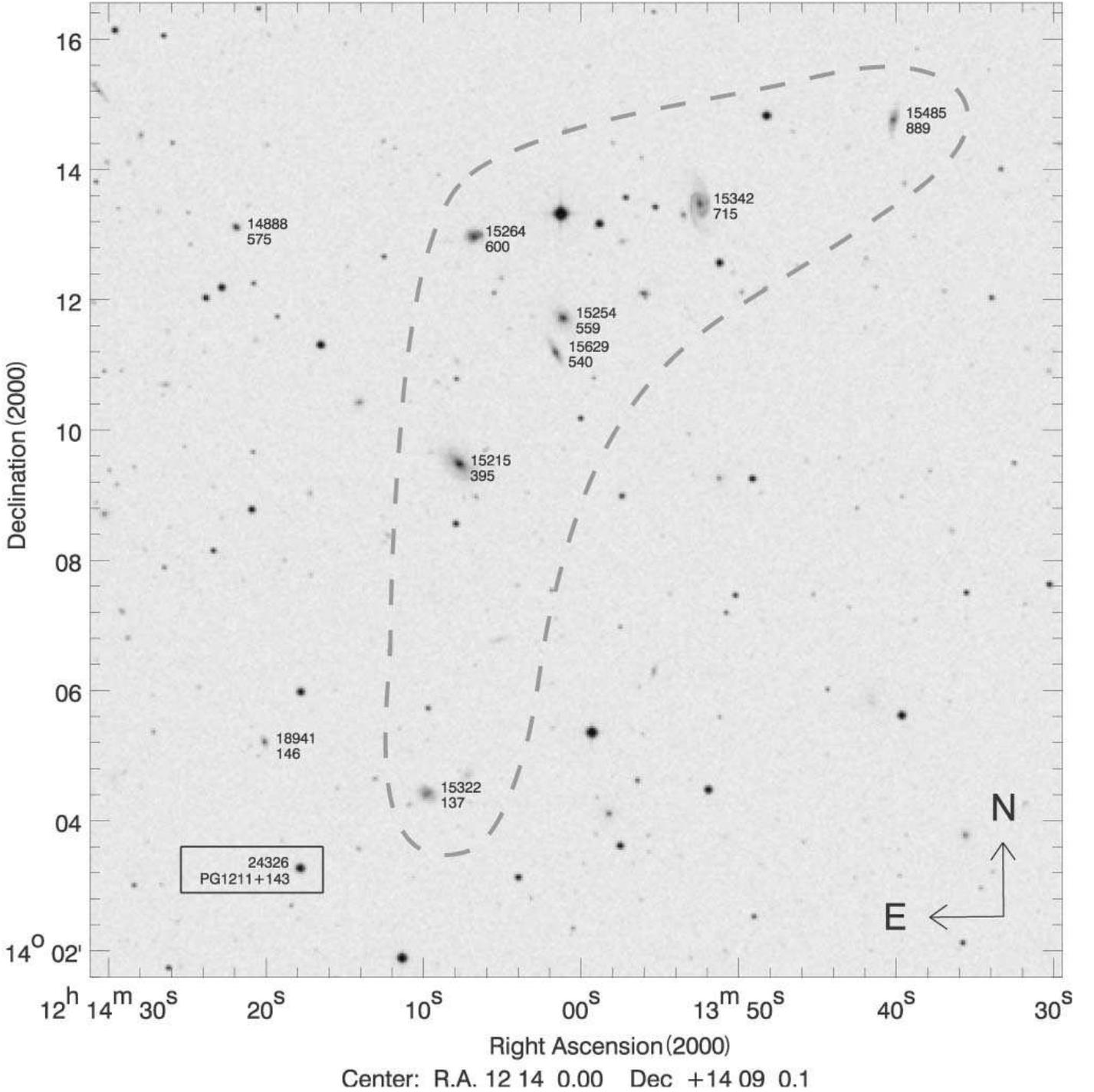}}} \caption{The
sky around PG~1211+143. This $15' \times 15'$ image from the POSS2
blue plates was obtained online from STScI. PG~1211+143 lies at
lower left. The galaxy group associated with the 15,300 \kms\
absorber complex extends to the northwest as marked. The isolated
galaxy associated with the 19,300 \kms\ absorbers lies $\sim 2'$
north of the QSO. The galaxies are labeled with their heliocentric
velocity ($cz_{\rm gal}$ in \kms) and their projected distance
($h_{70}^{-1}$ kpc) to the QSO sightline at $cz_{\rm gal}$.  These
distances assume a flat universe with $H_0 = 70$ km s$^{-1}$
Mpc$^{-1}$, $\Omega_{m} = 0.3$, and $\Omega_{\Lambda} = 0.7$. In
an attempt to determine whether the 15,300 \kms\ galaxy group
and/or filament extends SE beyond PG~1211+143, the galaxy field
was surveyed 3 - 5$'$ to the S and E of the QSO. We obtained 8
redshifts in the quadrants S and E of the displayed field and made
no additional discoveries of galaxies at the same redshifts as the
absorbers.  \label{pgmap}}
\end{figure*}

The galaxy field is pictured in Figure~\ref{pgmap}, and the galaxies
lying within $|\Delta v| = 500$ \kms\ of the two absorber complexes
are listed in Table~\ref{galtable}. They range in projected
separation from $\rho_{\perp} = 137h_{70}^{-1}$ to 889$h_{70}^{-1}$
kpc away from the QSO sightline (for Hubble constant $H_0 =
70h_{70}$ km s$^{-1}$ Mpc$^{-1}$). The seven CfA galaxies lie
northwest of PG~1211+143 in a spiral-rich group $\sim
750h_{70}^{-1}$ kpc in projected extent with velocity dispersion
$\sigma_{\rm cz} = 150$ \kms\ (if the nearby galaxy at $cz = 14,888$
\kms\ is included, $\sigma_{\rm cz}$ = 215 \kms).  We have extended
the Stocke et al. (2005) search for associated faint galaxies to the
east and south of PG~1211+143 but have not discovered any additional
galaxies at 15,300 or 19,300 \kms. Thus, the associated galaxies
appear to be confined to the northwest of PG~1211+143.  In \S~4 we
consider the possible connections between these galaxies, the group,
and the absorbers in the PG~1211+143 sightline.

\section{PHYSICAL CONDITIONS AND METALLICITY} \label{modelsection}

In this section, we construct models for the PG~1211+143
absorption-line systems using a combination of photoionization
equilibrium (PIE) and collisional ionization equilibrium (CIE)
models. We break down the observed absorption profiles into
individual velocity components, which are fitted separately but
simultaneously. In some cases we allow these fits to be guided by
expectations from modeling. We then regroup these components
according to their common characteristics (e.g., ionization stage,
linewidth, etc.) and analyze the groups as single systems. The
grouped component systems are labeled in Tables~\ref{15table} and
\ref{19table} and include column-density limits on additional
species when they constrain the models.

The proper interpretation of these complex absorption-line systems
proceeds here in a three-step process. In the first step, we use
the measured column densities, limits, linewidths, and relative
velocities to diagnose the physical conditions indicated by the
observed absorption. For our study, this means distinguishing PIE
gas from CIE gas, and determining the physical conditions for
each.  The second step is the phenomenological step, in which we
place these underlying physical processes in their proper context.
For example, hot \ion{O}{6}-bearing gas can arise in shocks,
turbulent mixing layers, or conductive interfaces -- physically
distinct origins that produce \ion{O}{6} with the same underlying
physical process (ionization by thermal electron collisions). The
third step in the interpretation is the selection of the
phenomenological explanation that best fits the available
information into a single physical picture. For example, we might
argue that the \ion{O}{6} arises in material ejected from the
nearby galaxies and shocked against photoionized Ly$\alpha$ forest
gas in the galaxy's extended gaseous halo. The first two steps are
carried out in this section.  In \S~4, we generalize the results
derived here and discuss the possible causes of the \ion{O}{6}
systems, including Galactic infall and outflow and hot-IGM models.

We model these absorbers with single-phase PIE and CIE models constructed
using the CLOUDY code (version 94.00; Ferland et al.~1998).  For PIE,
we assume that the gas is optically thin to an extragalactic ionizing
background with spectral shape from Shull et al.~(1999).  We normalize
this extragalactic ionizing background to a specific intensity $J_\nu =
10^{-23} J_{-23}$ erg~cm$^{-2}$~s$^{-1}$~Hz$^{-1}$~sr$^{-1}$ at 1 ryd.
The measured value of $J_{-23}$ is thought to be of order unity at $z
\approx 0$ (Shull et al. 1999; Tumlinson et al. 1999).  With a fixed
radiation field, a variation in the photoionization parameter ($U \propto
J_\nu / n_H$) represents a variation in the density $n_H$ (cm$^{-3}$).
Thus, for a given range in $U$, we state a corresponding range in density
and line-of-sight extent for the gas, assuming $J_{-23} = 1$, with the
caveat that these quantities are directly and inversely proportional
to the actual value of $J_{-23}$.  We also vary the metallicity ($Z$)
by scaling all heavy-element abundances relative to their solar values
with a uniform scale factor.  We then constrain the properties of the
models to be compatible with all absorption-line measurements and limits
associated with that velocity component. The quoted temperatures for model
clouds are derived self-consistently within CLOUDY for each value of $U$
and $Z$. For CIE, we use CLOUDY to calculate ion ratios as a function of
temperature, ignoring radiation. We assume the updated solar abundances
of carbon (C/H = $2.45 \times 10^{-4}$; Allende Prieto et al. 2001a)
and oxygen (O/H $= 4.90 \times 10^{-4}$; Allende Prieto et al. 2001b).

The conclusions we draw about the physical origins of these
absorbers contain the implicit assumption that the gas is in either
PIE or CIE. Because we consider only these mechanisms, the
uncertainties in the physical state of the detected gas correspond
only to the formal uncertainties in these models. There are other
ionization mechanisms that may be able to produce the detected
absorption, but which are much more difficult to constrain with the
available information. One such mechanism is non-equilibrium
collisional ionization (Shull \& McKee 1979; Shapiro \& Moore 1976),
in which the gas possesses either enhanced high- or low-ionization
states for its temperature. Such a model is difficult to constrain,
and difficult to pursue for these absorbers. We cannot know whether
the non-equilibrium gas was shocked to higher ionization, with a
transient excess of low ions, or has recombined from a higher
ionization state, with a transient excess of high ions. With such
ambiguity we would not be able to connect the physical state of the
gas to explanations for its origins, as we do for PIE and CIE in
\S~4. Also, the transient states associated with non-equilibrium
collisional ionization are brief, so it is unlikely that all these
absorbers would lie in such a state. We consider non-equilibrium
ionization to be a possibility, but one which cannot be suitably
tested against the data.

\subsection{Physical Conditions in the 15,300 km s$^{-1}$ System}

The absorption-line system at 15,300 \kms\ shows a remarkably
complex structure for a Ly$\alpha$ cloud. With profile fitting
informed by reasonable inferences from photoionization models, we
have identified eight distinct absorption components associated
with the broad Ly$\alpha$ (1130 m\AA\ total equivalent width;
Penton et al. 2004), six of which show metal-line absorption.
These appear in Table~\ref{15table} as A, B, B*, C, C*, and the
\ion{O}{6} component (the asterisks denote that the component has
the same velocity but higher ionization than its counterpart).
Blending is severe in components B, B$^*$, C, and C$^*$, so we
often fit more than one of them as a single component. Component D
(15,407 km s$^{-1}$) appears in \ion{H}{1} only; the apparent
\ion{Si}{3} $\lambda$1206 in component D is a Ly$\alpha$
interloper at $cz =$ 13,035 \kms\ that has been confirmed by a
detection of Ly$\beta$ in the {\em FUSE} data.  The column density
limits on metal lines do not provide a meaningful metallicity
limit for this system. The weak component E (15,570 km s$^{-1}$)
appears only in Ly$\alpha$ and has $N$(\ion{H}{1}) too low for a
meaningful metallicity limit to be determined. These
Ly$\alpha$-only components will not be discussed any further.  We
note that the Ly$\delta$ line at 15,300 \kms\ is blended with the
Ly$\epsilon$ line from 19,300 \kms. Still higher Lyman series
lines are too weak to give useful information or are inaccessible
due to blending, which is severe in the {\em FUSE} SiC2 channel
below 1000 \AA. The metallicity results for these components are
summarized in Table~\ref{metaltable}.

{\bf A (15,288 km~s$^{-1}$):} This component shows \ion{H}{1},
\ion{C}{3}, and \ion{C}{4}.  The \ion{H}{1} at A is blended with the
components B through C* in the Ly$\alpha$ and Ly$\beta$ transitions,
but in Ly$\gamma$ it is possible to separate components A and B -
C$^{*}$. The \ion{H}{1} and \ion{C}{3} in this component present
special difficulties in fitting.  The fitted doppler $b$ from
\ion{C}{4} was used to fix the $b$ for \ion{C}{3}, which is poorly
constrained on its own, owing to its location in the damping wing of
Galactic Ly$\beta$. If the column density for \ion{H}{1} is allowed
to float, the best fit is obtained for $b_{HI} = 6$ \kms, which
implies $T \leq 2200 K$, inconsistent with PIE models that produce
\ion{C}{4} at $T \sim$ 30,000 K and CIE models that produce
\ion{C}{4} at $T \sim 10^5$ K. We therefore adopt column densities
fitted with the assumption that $b_{HI} = 22$ \kms, or the thermal
broadening corresponding to 30,000 K for \ion{H}{1} (this is also
the median value for Ly$\alpha$ absorbers; Dav\'{e} \& Tripp 2001).
This assumption results in narrow error bars on the column density,
and it has only a modest effect ($\lesssim $ 0.1 dex) on the fitted
column density of the \ion{H}{1} associated with the B, B$^{*}$, C,
and C$^{*}$ components in Ly$\gamma$.

\begin{figure*}[t]
\centerline{\epsfxsize=1.0\hsize{\epsfbox{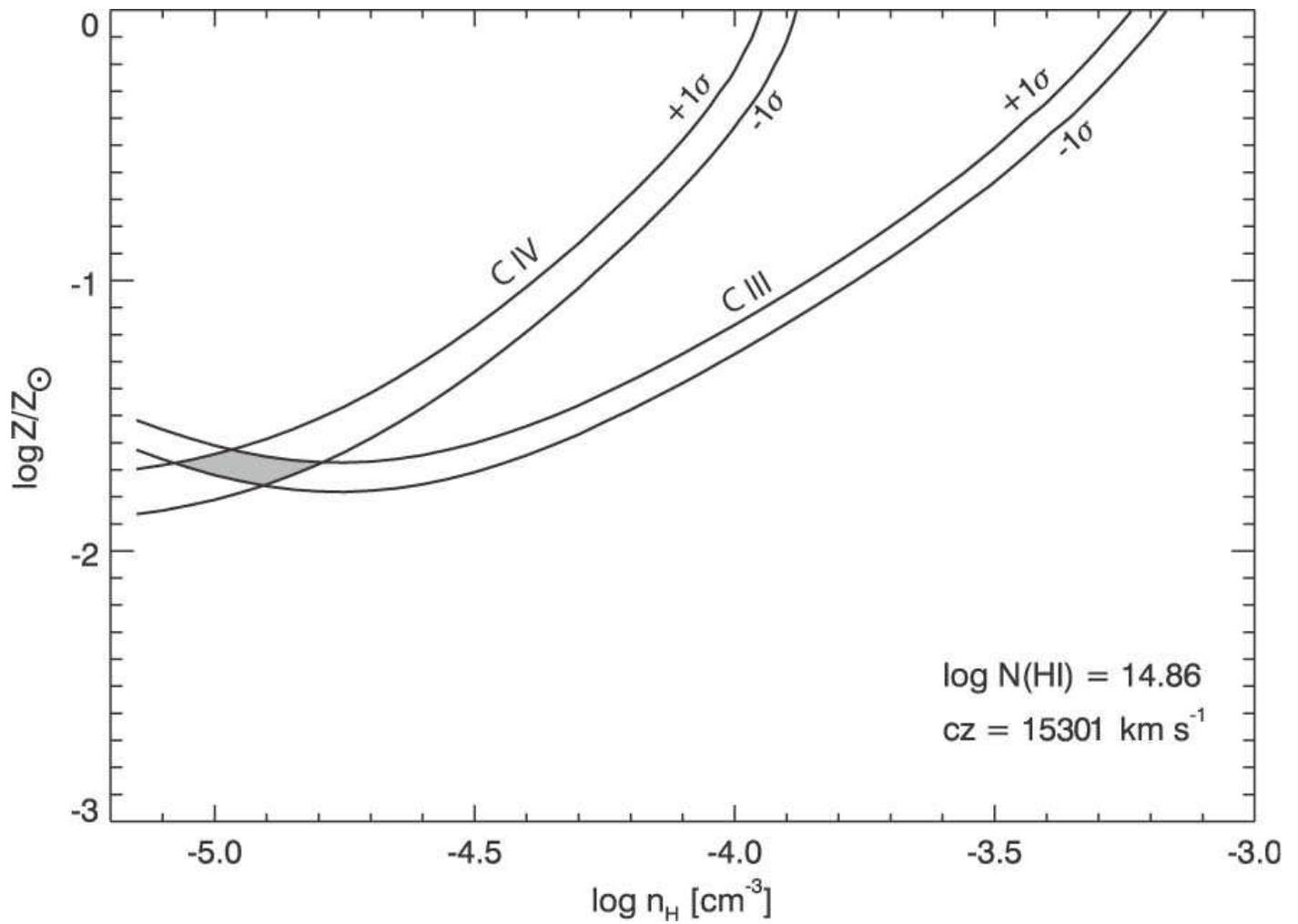}}} \caption{The
density-metallicity parameter space for the A component (15,301
\kms) of the 15,300 \kms\ system. The permitted range of parameters
is marked with the shaded region, where the $\pm 1 \sigma$ ranges of
permitted column density for C~III and C~IV
overlap.\label{photoifig}}
\end{figure*}

The $1\sigma$ ranges on the column densities of \ion{C}{3} and
\ion{C}{4} constrain a single-phase PIE model to have ionization
parameter $-1.6 \leq \log U \leq -1.5$.  With $\log N$(\ion{H}{1}) $
= 14.86 \pm 0.10$ set by Ly$\gamma$ (the only profile where
component A can be separately fitted), the metallicity of the gas is
constrained to be $\log Z/Z_{\odot} \simeq -1.7 \pm 0.1$ (2\%
solar). The ionization parameter corresponds to a density $\log n_H
\simeq -5.0 \pm 0.2$, pressure $P/k = 2.24n_HT \simeq 0.67$
cm$^{-3}$ K, and line-of-sight extent $d = 180 - 400$ kpc. The
corresponding range in temperature $T = 35,000 - 38,000$ K is
marginally consistent with our assumption of $b_{HI} = 22$ \kms. We
note that the small formal uncertainties in these parameters are due
in large part to this assumption about the fitted $b_{HI}$. We use
this component to illustrate the photoionization models we use
throughout this section. Figure~\ref{photoifig} shows the permitted
$\pm 1\sigma$ ranges of \ion{C}{3} and \ion{C}{4} column density,
which overlap in the shaded region.

{\bf C~IV (15,319 \kms):} We also note the additional \ion{C}{4}
component at 15,319 \kms, which does not appear to match any other
metal-line or \ion{H}{1} velocity component in the system, and
therefore has not received its own label. If this \ion{C}{4}
component arises in gas with $\log N(HI) \simeq 14.5$, $\log U
\simeq -1.8$, and $Z/Z_{\odot} \simeq 0.1$, it is too weak to appear
in any other species covered by our data. This PIE model predicts
$\log N$(\ion{N}{5}) $ = 12.3$, $\log N$(\ion{C}{3}) $= 12.7$, and
$\log N$(\ion{O}{6}) $ = 13.1$, below the levels that would be
visible in the observed metal-line profiles. Because this ionization
parameter and metallicity are roughly typical of the other
components at 15,300 \kms, it is quite likely that this \ion{C}{4}
component contributes little to the observed metal-line profiles. If
we include an additional component with $\log N$(\ion{H}{1}) = 14.5
and $b = 10 - 20$ \kms\ in the fit to Ly$\gamma$, we find that it
has minimal impact, within the 1$\sigma$ errors, on every component
but the highly uncertain component A. The quoted error for component
A has therefore been increased from 0.05 to 0.10 dex to compensate.
We have no other information about this component and cannot analyze
it any further.

{\bf B and B$^{*}$ (15,340 km~s$^{-1}$):} This velocity component
appears in \ion{C}{3}, \ion{Si}{3}, \ion{C}{4}, and \ion{Si}{4}.
Linewidths of $b \sim 20$ \kms\ for both \ion{C}{3} and \ion{Si}{3}
appear to distinguish the moderately-ionized component B from the
highly ionized component B$^{*}$ seen in \ion{C}{4} and \ion{Si}{4},
which has $b \sim 7$ \kms. This disparity in linewidths is supported
by the fact that it appears in data from both {\em FUSE}
(\ion{C}{3}) and STIS (\ion{Si}{3}, \ion{Si}{4}, and \ion{C}{4}),
with higher resolution. However, it may be that the saturated
\ion{C}{3} and strong \ion{Si}{3} conceal additional weak, narrow
components that align with the high ions. This possibility
illustrates one reason why we cannot produce detailed
photoionization models for these blended components. This issue is
treated in more detail below during the discussion of all components
B through C$^*$.

{\bf C and C$^{*}$ (15,354 km~s$^{-1}$):} This velocity component
appears in \ion{C}{2}, \ion{Si}{2}, \ion{Si}{3}, \ion{C}{4}, and
\ion{Si}{4}. These low- and high-ionization species have narrow
linewidths and the same velocity, but we group them separately
because they are not usually produced together in simple PIE models.
For a grid of solar-metallicity models covering $-4 \leq \log U \leq
-1$ and $\log N$(\ion{H}{1}) $ = 14.0 - 16.0$, we find that there is
no single region of parameter space ($U$, $N$) that matches the
observed column density of these five ions simultaneously. The ion
pairs \ion{C}{4}-\ion{Si}{4} and \ion{C}{2}-\ion{Si}{2} appear
together more commonly than the element pairs \ion{C}{2}-\ion{C}{4}
and \ion{Si}{2}-\ion{Si}{4}. For this reason, we group the high ions
and low ions together in different pairs.  We define component C to
include \ion{Si}{2} and \ion{C}{2}, and component C$^{*}$ to include
\ion{C}{4} and \ion{Si}{4}. The narrow component in \ion{Si}{3} at
15,356 \kms\ could correspond to either one, but we place it with
C$^{*}$, consistent with most PIE models in which \ion{Si}{3} goes
with \ion{Si}{4}. A narrow component forced into the \ion{C}{3}
profile at this position does not substantially affect the fits for
A and B in \ion{C}{3}, but this additional \ion{C}{3} component
cannot be well-constrained in column density and linewidth. Since we
do not attempt detailed models of these components, their exact
groupings are not critical to our conclusions.

{\bf B, B$^*$, C, C$^{*}$ combined (15,340 - 15,354 km~s$^{-1}$):}
These four components and the 15,319 \kms\ \ion{C}{4} component
probably all contribute to the observed Ly$\alpha$ profile at 15,349
\kms\ and the broad \ion{H}{1} absorption in Ly$\beta$ and
Ly$\gamma$. Because these apparent components arise so closely
together in velocity space, we now consider them all in relation to
one another. Deriving detailed photoionization models for these
blended and ambiguous components would require better knowledge of
how to apportion the \ion{H}{1} than we can derive from fits to the
observed profiles. Because of this ambiguity, we have not attempted
to match detailed photoionization models to the individual
components. Instead, we seek three more modest goals: (1) to
establish that photoionization is the likely production mechanism
for the observed ion species; (2) to obtain reasonable estimates of
or limits on the metallicity in one or more of the components; and
(3) to determine the extent to which the observed \ion{O}{6} is
associated with the gas traced by the low ions.

First, we assess the viability of CIE models for the moderately
ionized gas seen in components B, B*, C, and C*.  For B* and C*, we
use the observed \ion{C}{4}/\ion{Si}{4} ratio to estimate $\log T
\simeq 4.9 - 5.0$. In both B* and C*, the predicted and observed
silicon linewidths are consistent ($b_{th}^{Si} \sim 7$ \kms), but
the implied thermal linewidth for carbon ($b_{th}^C \sim 11 - 12$
\kms) is roughly twice the fitted value, and the implied $b_{th}^H =
40$ \kms\ thermal linewidth is too broad to fit the Ly$\gamma$
profile\footnote{For single-phase gas at temperature $T$, $b^2 =
(2kT/m) + b_{nt}^2$, where $m$ is the atomic mass and $b_{nt}$ is
the turbulent and/or non-thermal velocity dispersion.}. We conclude
that CIE is not viable for B* and C*.  For component A, the
one-to-one ratio of \ion{C}{3} to \ion{C}{4} implies $\log T \simeq
5.0$, which predicts an \ion{H}{1} linewidth $b_{th}^H \simeq 41$
\kms, too broad for the observed Ly$\gamma$ profile. For the narrow
C component we derive $\log T \simeq 4.2$ from the
\ion{C}{2}/\ion{Si}{2} ratio, but the implied \ion{H}{1} column at
this temperature is $\log N$(\ion{H}{1}) $ \simeq 16.5$, too high to
be accommodated by the Lyman series profiles. In the case of the
broad component B, CIE is viable at $\log T \simeq 4.8$, where
$b_{th}^C \simeq 9$ \kms\ and $b_{th}^{Si} \simeq 6$ \kms . These C
and Si linewidths are consistent with the observed values if $b_{nt}
\simeq 18$ \kms, but the implied \ion{H}{1} linewidth of $b_{th}^H
\simeq 33$ \kms\ violates the observed linewidth constraint on
\ion{H}{1} ($b_{HI} = 6 - 22$ \kms\ fitted from Ly$\gamma$).

Since CIE has been eliminated as the source of ionization in the
moderately ionized components B - C*, we next turn to
photoionization models to judge their ability to produce the
observed ionization. A grid of photoionization models ranging over
$-4 \leq \log U \leq -1$ and $\log N$(\ion{H}{1}) $ = 14.0 - 16.0$
with solar metallicity shows that the observed column densities of
\ion{C}{2}-\ion{C}{4}, \ion{Si}{2}-\ion{Si}{4} and their relative
ratios can generally be produced for $\log N$(\ion{H}{1}) $ \simeq
14.5 - 15.0$, such that four components could be accommodated within
the Ly$\gamma$ profile. This means, however, that their
metallicities, on average, cannot lie far below solar without
violating the total \ion{H}{1} constraint. This limit is especially
severe for the \ion{Si}{2} in component C, which requires $\log
N$(\ion{H}{1}) $ \gtrsim 14.9$ at solar metallicity and higher
$N$(H~I) at sub-solar metallicity. This component must have $Z
\gtrsim $ 40\% solar, even if we assign it all the available
\ion{H}{1}. Thus, at least one of the components with
low-to-moderate ionization is expected to have a relatively high
metallicity for diffuse IGM gas. In \S~4 we interpret this as a clue
to the absorbers' physical origins.

{\bf \ion{O}{6} (15,341 km~s$^{-1}$):} The strong \ion{O}{6} line
provides a well-measured column density $\log N$(\ion{O}{6}) $ =
14.26^{+0.05}_{-0.08}$.  It is unclear how much \ion{H}{1} is
associated with this \ion{O}{6}.  If it is photoionized, we can
infer an acceptable range in metallicity and ionization parameter
for an assumed $\log N$(\ion{H}{1}) and judge whether the model is
incompatible with the other absorption lines.  Given the observed
\ion{H}{1} profiles, a column density of $\log N$(\ion{H}{1}) $
\simeq 14.5$ for photoionized \ion{O}{6} is assumed.  Under this
assumption we derive $-1.6 \leq \log U \leq -0.9$, $d = 46 - 2000$
kpc, $-1.3 \leq \log Z/Z_{\odot} \leq 0$, and $T = 14,000-40,000$ K.
These models have $\log N$(\ion{C}{4}) $ = 13.1 - 14.3$, $\log
N$(\ion{C}{3}) $ = 12.5 - 14.1$, $\log N$(\ion{N}{5}) $ = 13.1 -
13.7$, and $\log N$(\ion{Si}{4}) $ = 8.5 - 12.2$.  At the lower end
of these column density ranges, and with $b \simeq 53$ \kms\ for all
species (dominated by the nonthermal broadening shown by
\ion{O}{6}), these lines (especially \ion{C}{4}) are very difficult
to exclude based on the observed profiles. However, such a model
cannot readily explain the discrepancy between the predicted thermal
linewidth for \ion{O}{6} ($b_{th}^O \leq 6.4$ \kms\ at $T \leq$
40,000 K) and the observed value of $b_{OVI} = 53$ \kms, which would
need to be almost completely nonthermal. At the larger sizes, line
broadening by the Hubble flow over 2 Mpc would exceed the observed
linewidth. Although we consider this model unlikely, owing to this
linewidth discrepancy, we cannot exclude it on the basis of the
predicted column densities alone. (See \S~\ref{conclusionsection}
for a discussion of how photoionized and collisionally ionized
\ion{O}{6} can be distinguished by X-ray observations.)

%Initially, we assume $\log N$(\ion{H}{1})
%$ = 15.0$, adopt $J_{23} = 1$, and limit the extent of the gas to $\leq 2$
%Mpc to further constrain the range of ionization parameters.  This model
%gives $-1.7 \leq \log U \leq -1.2$, $d = 112 - 2000$ kpc, $-1.3 \leq \log
%Z/Z_{\odot} \leq -0.3$, and $T = 18,000-35,000$ K.  Over the same range
%in ionization parameter, $\log N$(\ion{C}{4}) $ = 13.7 - 14.5$, $\log
%N$(\ion{C}{3}) $ = 13.5 - 14.4$, $\log N$(\ion{N}{5}) $ = 13.5 - 13.8$,
%and $\log N$(\ion{Si}{4}) $ = 10.0  - 12.4$. These lines, especially
%the strong \ion{C}{4}, would appear in the observed profiles. Their
%absence is an indication that the \ion{O}{6} is not photoionized with
%$\log N$(\ion{H}{1}) = 15.0.

If the \ion{O}{6} is collisionally ionized, its linewidth
corresponds to $2.8 \times 10^6$ K (with a $1\sigma$ upper limit of
$T \leq 4 \times 10^6$ K at $b = 63$ \kms), a temperature at which
no other species covered by our data would appear (this model
predicts $\log N$(\ion{H}{1}) $ \simeq 13.0$). At the maximum
ionization fraction of \ion{O}{6} ($f_{O VI} = 0.22$ at $\log T \sim
5.45$), the predicted thermal linewidth for oxygen is $b_{th}^O =
17$ \kms, requiring $b^{O}_{nt} = 50$ \kms\ to achieve the observed
profile. An unconstrained portion of this additional linewidth may
be velocity shear behind a bow shock. The contribution of shear lies
in the range $\Delta V = (1-2) V_s \sin \theta$ \kms, where $V_s$ is
the shock velocity and $\theta$ is the angle between the infalling
material and the shock vector. This shear contribution can be
significant for shocks strong enough to generate \ion{O}{6} ($V_s
\gtrsim 130$ \kms), but $\theta$ is unconstrained. Given these
uncertainties and the ambiguous apportionment of \ion{H}{1} into the
various photoionized components at the same velocity, we cannot
derive the same robust temperature or metallicity limits for the
\ion{O}{6} here as we do for the 19,300 \kms\ system (see \S~3.2).
We consider collisional ionization to be a more likely mechanism
than photoionization for the \ion{O}{6}, but the uncertainties in
the nearby blended photoionized clouds preclude more definite
conclusions.

\subsection{Physical Conditions in the 19,300 km s$^{-1}$ System}

 This system consists of five components of \ion{H}{1}
(total Ly$\alpha$ profile equivalent width 880 m\AA; Penton et al.
2004) and moderate ionization species (\ion{C}{3} and \ion{Si}{3})
that appear to be photoionized (A, B, D). In addition, there are
two strong \ion{O}{6} components with unclear alignment or
association with the \ion{H}{1} components. The E component
appears in Ly$\alpha$ only and will not be discussed any further.
We note that the Ly$\epsilon$ line at 19,300 \kms\ is blended with
the Ly$\delta$ line from 15,300 \kms. Still higher Lyman series
lines are too weak to give useful information or inaccessible due
to severe blending in the {\em FUSE} SiC2 channel below 1000 \AA.
The metallicity results for these components are summarized in
Table~\ref{metaltable}.

{\bf A (19,274 km s$^{-1}$):} This component is highly uncertain in
both $N$(\ion{H}{1}) and $N$(\ion{C}{3}). The only distinct
measurement of N(\ion{H}{1}) comes from the A component of
Ly$\gamma$, which exhibits a relatively narrow linewidth that is
difficult to constrain. In the profile fits, we required that
$b_{HI} \geq 10$ \kms, which corresponds roughly to ionized gas at
$T \gtrsim 8000$ K. Below this temperature H will be only partially
ionized and \ion{C}{3} will not appear. This requirement leaves only
a narrow range of permitted $N$ and $b$ for \ion{H}{1} (see
Table~\ref{19table}). The \ion{C}{3} column density is also poorly
constrained. However, we are still able to produce reasonable
photoionization models.  For the combined \ion{C}{3} measurement and
the \ion{C}{4} limit, we derive an ionization parameter range of
$-2.8 \leq \log U \leq -2.5$.  The corresponding range in
temperature $T$ is 8,700 $\leq T \leq$ 16,300 K.  We adopt
$N$(\ion{H}{1}) $ = 14.46^{+0.12}_{-0.05}$ from the fit to
Ly$\gamma$, which is less saturated and blended than Ly$\alpha$ and
Ly$\beta$.  The acceptable range in metallicity is $-0.4 \leq \log
Z/Z_{\odot} \leq 0.0$.  For $J_{-23} = 1$, the range in ionization
parameter corresponds to a line-of-sight extent of $0.14-1.4$ kpc.
If we include the measured \ion{Si}{3} and the limits on \ion{Si}{2}
and \ion{Si}{4}, the allowed parameter ranges narrow further to
$-2.7 \leq \log U \leq -2.6$, $-0.2 \leq \log Z/Z_{\odot} \leq 0.0$,
$-3.9 \leq \log n_H \leq -3.8$, $T = 9,500 - 12,900$ K and
line-of-sight extent $d = 170 - 400$ pc.  This absorber shows that
additional species can constrain a photoionization model better than
tighter limits on fewer species. Despite the large uncertainties in
the profile fits, we believe that this component arises in warm,
moderately photoionized gas with $\gtrsim 40$\% solar
metallicity\footnote{The integrated apparent column density
measurement of $N$(C~III), as described in \S~4.5, restricts this
model to $\log (Z/Z_{\odot}) \simeq -0.1$.}.

{\bf B (19,317 km s$^{-1}$):} For the observed \ion{C}{3} column
density measurement and \ion{Si}{2} and \ion{C}{4} limits, we find
that this component is well-fitted by photoionization with $-3.0
\leq \log U \leq -2.4$.  The corresponding range in temperature is
$T = 13,000 - 23,000$ K.  In this case, the constraints on the
metallicity and the ionization parameter are highly correlated, but
not narrowly constrained individually.  For $\log N$(\ion{H}{1}) $ =
15.09^{+0.06}_{-0.07}$ (from Ly$\gamma$, the only profile where A
and B can be separated), the acceptable range in metallicity is
$-1.3 \leq \log Z/Z_{\odot} \leq -0.6$.  For $J_{-23} = 1$, the
range in ionization parameter corresponds to an extent of $0.23 -7$
kpc. This component appears to arise in warm photoionized gas but
with less well-constrained properties than component A.

{\bf D (19,474 km s$^{-1}$):} For the observed \ion{C}{3} column
density we find $-2.3 \leq \log U \leq -1.0$.  The corresponding
range in temperature $T$ is $10,600 \leq T \leq 20,700$ K.  For
$\log N$(\ion{H}{1}) $ = 13.65$, from Ly$\beta$, the acceptable
range in metallicity is $-0.2 \leq \log Z/Z_{\odot} \leq 0.0$.  For
$J_{-23} = 1$, the range in ionization parameter corresponds to an
extent of $0.17 - 85$ kpc. When combined with the upper limit $\log
N$(\ion{C}{4}) $ \leq 12.5$, this model yields ionization parameter
$\log U \approx -2.2$, and metallicity $\log Z/Z_{\odot} \approx
-0.1$, as well as an extent of $\simeq 300$ pc and a temperature $T
\approx 11,000$ K.  This component arises in warm, moderately
photoionized gas with $\gtrsim 40$\% solar metallicity. A model
based on the lower $\log N$(\ion{H}{1}) = $13.40$ (within the errors
on the Ly$\beta$ fit) from the Ly$\alpha$ fit requires a
correspondingly higher metallicity.

{\bf \ion{O}{6} (19,356 and 19,434 km s$^{-1}$):} In addition to the
three photoionized (\ion{C}{3}- and \ion{Si}{3}-bearing) components,
this system exhibits two strong, broad \ion{O}{6} components. The
\ion{O}{6} component at 19,356 \kms\ lies well away from any clearly
defined \ion{H}{1} component, but the \ion{O}{6} line at 19,434
\kms\ may correspond to Ly$\alpha$ component C. Here we discuss CIE
models for both absorbers and PIE models for the second \ion{O}{6}.

First, we assume that the \ion{O}{6} line at 19,434 \kms\ arises
in the same single-phase photoionized gas as the Ly$\alpha$
component at 19,423 \kms\ (C), as suggested by their close
velocity coincidence and compatible linewidths.  Under this
assumption, we derive a range of ionization parameter $-0.9 \leq
\log U \leq 0.0$, density $-6.5 \leq \log n_H \leq -5.6$, and
temperature $T = 42,000 - 77,000$ K from the $N$(\ion{O}{6})
measurement and $N$(\ion{C}{4}) limit. With $N$(\ion{H}{1}) =
13.65, this model gives metallicity $-1.7 \leq \log (Z/Z_{\odot})
\leq -1.3$ and line-of-sight extent $d = 400$ kpc to 44 Mpc. These
larger sizes are unreasonable, but there is a narrow range of
parameter space with $d \sim 400$ kpc and $\sim 5$\% solar
metallicity that cannot be conclusively excluded.  However, in
this region, the model temperatures are $T \sim 40,000$ K, which
is only marginally consistent with the observed \ion{H}{1} and
\ion{O}{6} linewidths For $T = 42,000$ K and $b_{nt} = 10$ \kms,
we derive $b_{HI} = 28$ \kms\ and $b_{OVI} = 11$ \kms, much
narrower than the observed $b_{OVI}$; for larger cloud sizes the
Hubble flow would broaden the line beyond the observed width. This
last factor suggests that PIE is only a marginal fit to the
available constraints. It is also rather unlikely that a region as
large as 400 kpc would have been enriched uniformly to 5\% solar
metallicity.  The \ion{O}{6} appears collisionally ionized, since
$b_{th}^{O} = 17$ \kms\ at $T = 10^{5.45}$ K, but we cannot
completely reject PIE models for this \ion{O}{6}.  As for the
15,300 \kms\ system, the 19,434 \kms\ \ion{O}{6} line could be
photoionized if it is extremely large, metal-enriched, and
substantially broadened by non-thermal motions.

However, the absence of associated \ion{C}{4}, clear-cut \ion{H}{1},
and the broad linewidths of both \ion{O}{6} absorbers suggest that
they are produced by CIE at $\log T \simeq 5.5$. We adopt this
hypothesis and use the observed \ion{O}{6} column densities, the
\ion{C}{4} limit, and the line profiles to constrain the temperature
and metallicity of the ionized gas. For this purpose we use a
conservative limit \ion{C}{4}, $\log N$(\ion{C}{4}) $<13.1$, derived
by integrating noise in the continuum over a 100 \kms\ range for
each \ion{O}{6} line (19,300 - 19,400 \kms\ and 19,400 - 19,500
\kms). This limit corresponds to an unseen \ion{C}{4} line assumed
to arise in the same gas as \ion{O}{6}, and therefore roughly as
broad. For CIE and a solar C/O ratio, \ion{O}{6} achieves a higher
column density than \ion{C}{4} at $\log T > 5.3$. The observed
\ion{O}{6} columns and \ion{C}{4} limit (Table~\ref{19table})
combine to constrain the temperature to $\log T \geq 5.36$ for the
$cz = 19,356$ \kms\ absorber and to $\log T \geq 5.31$ for the $cz =
19,434$ \kms\ absorber. These temperatures correspond to thermal
linewidths of $b^{O}_{th} = 15$ \kms\ for \ion{O}{6} and $b^{H}_{th}
= 59 - 62$ \kms\ for \ion{H}{1}. The observed linewidths imply
non-thermal broadening of $b_{nt} \leq 34$ \kms\ and $b_{nt} \leq
15$ \kms\ for the two components, respectively. Even if nonthermal
broadening is ignored, these thermal linewidths are too broad to be
consistent with the Lyman lines. This discrepancy suggests that if
there is any significant \ion{H}{1} associated with these
collisionally ionized \ion{O}{6} lines, it is not carried by any of
the identified components A - E.

With limits on the temperature and linewidths of the observed
\ion{O}{6}, we can estimate the amount of associated \ion{H}{1}
present if we assume that the \ion{O}{6} arises at $\log T \simeq
5.45$, where it appears in CIE at its maximum ratio to \ion{H}{1},
$\log \left[f_{OVI}/f_{HI}\right] = 5.13$. Off this peak, at higher
and lower $T$, a given observed \ion{O}{6} column density will
necessarily correspond to more \ion{H}{1}. At this maximum value we
can use the observed \lya\ profiles to constrain the oxygen
abundance O/H, where O/H = $\frac{N(OVI)}{N(H~I)} \times
\frac{f_{HI}}{f_{OVI}}$.  We do this by adding to the profile fits
two \ion{H}{1} components at fixed $cz$ and $b$ corresponding to the
observed \ion{O}{6} components. We allow $N$(\ion{H}{1}) to vary to
obtain a new best fit to the \lya\ profile, and we obtain $\log
N$(\ion{H}{1}) = 13.45 at $cz= 19,356$ \kms\ and $\log
N$(\ion{H}{1}) $ = 13.44$ at $cz = 19,434$ \kms\ (see bottom panel
of Figure~\ref{stack19}). For the solar O/H, these \ion{H}{1}
columns correspond to lower limits on the O abundance of [O/H] $\geq
-1.19$ (6\% solar) and [O/H] $\geq -1.63$ (2\% solar), respectively.
These are lower limits because we have maximized $\log
[f_{OVI}/f_{HI}]$; off the peak at $\log T \simeq 5.45$, higher
[O/H] is necessary to match the observed \ion{O}{6} and still
accommodate it within the \lya\ profile. Thus, for CIE, we find that
these hot clouds have $T \gtrsim 250,000$ K and $\gtrsim 2 - 6$\%
solar metallicity (see \S~4.8 for an estimate of the IGM metallicity
at $\sim$150 kpc from galaxies.)

The broad linewidths of the detected \ion{O}{6} components suggest
that they are collisionally ionized at $\log T \gtrsim 5.4$. We now
explore the implications of this conclusion for the physical origins
of this hot gas. In general, the production mechanism for hot,
collisionally-ionized \ion{O}{6} in intergalactic space is not well
understood. Known systems appear to arise from a diverse set of
processes; among these are Galactic fountains, interfaces between
hot substrates and infalling clouds, shear instabilities, and
turbulent mixing layers. However, these processes have difficulty
producing the observed column densities $\log N$(\ion{O}{6}) $\sim
14$ and ratios to \ion{C}{4}, \ion{Si}{4}, and \ion{N}{5}
(Indebetouw \& Shull 2004a,b).

The most likely location for the collisionally ionized hot gas is
the post-shock region behind a shock front between two interacting
clouds (detectable \ion{O}{6} requires $\Delta v \gtrsim 150$ \kms\
and $Z \gtrsim 0.01 Z_{\odot}$). The physical conditions in the two
\ion{O}{6} components and the presence of other IGM absorbers close
in $\Delta v$ motivate this view.  The other detected species are
consistent with a model in which the \ion{O}{6} is produced in the
post-shock region of a cloud interacting with one of the warm
photoionized components. To validate this model, we assume that a
standard adiabatic shock achieves a post-shock temperature $T_{s} =
(1.36 \times 10^{5}$ K) (V$_s$/100 \kms)$^2$, where V$_s$ is the
shock velocity in \kms\ and $\mu_s = 0.6m_H$ is the mean molecular
weight of fully-ionized, post-shock gas with $n_{He}/n_{H}= 0.08$
(Shull \& McKee 1979; Draine \& McKee 1993). This relation and the
temperature limits in CIE, set by N(\ion{O}{6})/N(\ion{C}{4}),
determine which photoionized components (A - E) can trace the
pre-shock material. Table~\ref{19shocktable} lists the relative
velocities between the two \ion{O}{6} components and the
photoionized components. In each case, three components have
sufficient relative velocity to achieve the high post-shock
temperatures ($\log T \gtrsim 5.3$) implied by the
N(\ion{O}{6})/N(\ion{C}{4}) ratios. We also list the sound speed for
the pre-shock gas as derived from the PIE models for components A,
B, and D, and from the observed linewidth for E.  These sound speeds
and relative velocities imply strong shocks, with Mach numbers 7 -
19 and density enhancement approaching 4 behind the shock.  The
densities implied by the PIE models range from $\log n_H = -5.5$ to
$-3.2$. Together with the 2 - 6\% solar metallicity limits derived
for the \ion{O}{6}-bearing gas, these densities imply a radiative
cooling time given by:
\begin{equation}
t_{\rm cool} \lesssim \left( 10^8 \,  {\rm yr} \right)
\left(\frac{10^{-4} \, {\rm cm}^{-3}}{n_H} \right) \left(
\frac{0.03Z_{\odot}}{Z} \right).
\end{equation}
If the \ion{O}{6}-bearing gas does not have substantially higher
metallicity than the $2 - 6$\% lower limits implied by the line
profiles, then the post-shock gas can have cooling times of $>$ 1
Gyr at the lowest model densities ($\log n_H \sim -5.5$). At the
higher densities and 30\% metallicities, as in the photoionized
components, the cooling time can approach 1 Myr, making it unlikely
that absorbers of this type would be commonly seen. These lower
densities for the post-shock gas imply pathlengths that are too
long, favoring the higher densities permitted by the photoionization
models.

The \ion{O}{6}-bearing components in the 19,300 \kms\ system toward
PG~1211+143 exhibit two robust features that favor their origin in
collisionally ionized hot gas behind a strong shock front. First,
they have linewidths too broad to be consistent with the \ion{H}{1}
and low ions seen nearby in velocity. Second, they have higher
column densities than expected from some alternative models
(conductive interfaces, turbulent mixing layers, Indebetouw \& Shull
2004a). We favor a model in which the \ion{O}{6} arises behind
strong shock fronts between low-metallicity clouds and one of the
low-density photoionized components seen in \ion{H}{1} and
\ion{C}{3}. This scenario is the simplest explanation for the
observed column densities and line profiles that fits all of the
available information. However, this picture does not uniquely
constrain other important properties of the system.  For instance,
this picture does not depend on whether the \ion{O}{6}-bearing
shocked material is ejected from or falling toward the nearby galaxy
(146$h_{70}^{-1}$ kpc away at $|\Delta v| \sim 400$ \kms). In fact,
this model does not require a causal connection between the galaxy
and the absorbers. In \S~4, we describe how such a connection could
be established, and what it would mean for the IGM-galaxy
connection.

\section{PHYSICAL ORIGINS OF THE ABSORBERS } \label{originsection}

\subsection{Physical Models of the Absorbers}

In this section, we describe various physical models for the detected
absorption complexes. These models complete the three-step interpretations
begun in \S~3 by placing the physical conditions, ionization mechanisms,
and metallicities into the context of descriptive models.  We consider
galactic halo, high-velocity cloud, dwarf starburst wind, and intragroup
medium models for the absorbers.  We begin by comparing the PG~1211+143
systems with other \ion{O}{6} systems to see if trends emerge from the
existing data.

Ongoing efforts to detect \ion{O}{6} in the low-$z$ IGM with {\it
FUSE} have found 127 \lya\ absorbers ($W_{\lambda} \geq 80$~m\AA)
with acceptable \ion{O}{6} data, 48 of which are detected in
\ion{O}{6} at $\geq 4 \sigma$ significance (Danforth \& Shull 2004).
Thus, $\sim$40\% of all \lya\ absorbers contain both \ion{H}{1} and
\ion{O}{6} in the $N$(\ion{H}{1}) range for which high-quality FUSE
spectra exist. The ``multiphase ratio", N(\ion{H}{1})/N(\ion{O}{6}),
varies substantially, from $\sim 0.1$ to greater than 100 (Danforth
\& Shull 2004), perhaps indicating a wide range in shock velocities
and/or O/H abundances in these multiphase systems.

A smaller number of Ly$\alpha$--\ion{O}{6} absorbers have been
studied by both {\em FUSE} and STIS/E140M.  We prefer these for
detailed comparisons because they closely match the data quality,
line coverage, and information content of the spectra we present
here. Although the circumstances of the observations in these other
studies are similar to our own, and we restrict our comparisons to
only those systems that show \ion{O}{6}, their properties are quite
diverse. Here, we make detailed comparisons to the STIS and {\em
FUSE} studies of the sightlines H1821+643 (Tripp et al. 2001),
PKS~2155-304 (Shull, Tumlinson, \& Giroux 2003), PKS~0405-1219 (Chen
\& Prochaska 2000; Prochaska et al. 2004), 3C~273 (Sembach et al.
2001; Tripp et al. 2002), and PG~0953+415 (Savage et al. 2002).

\subsection{Comparisons to 15,300 \kms}

The system at 15,300 \kms\ toward PG~1211+143 bears more than a
superficial resemblance to the partial Lyman limit system (LLS)
toward PKS~0405-1219 as reported by Chen \& Prochaska (2000) and
later studied thoroughly by Prochaska et al. (2004). Both systems
show broad \ion{O}{6} offset in velocity from the strongest
\ion{H}{1} component, and both show clear multiphase structure
with a wide range of ionization conditions. With $\log
N$(\ion{H}{1}) $ = 16.45 \pm 0.05$, the PKS~0405-1219 system shows
\ion{C}{2}, \ion{C}{3}, \ion{N}{2}, \ion{Si}{2}, \ion{Si}{3},
\ion{Si}{4}, and \ion{N}{5} at column densities 0.3 - 1.0 dex
higher than we detect at 15,300 \kms\ (the STIS data presented by
Prochaska et al. 2004 did not cover \ion{C}{4}). This absorber has
roughly 1 dex more \ion{H}{1} than the PG~1211+143 15,300 \kms\
absorber. This trend continues through the low ions (\ion{Si}{2},
\ion{C}{2}) and the moderate-to-high ions where they are available
in both cases (\ion{Si}{3}, \ion{Si}{4}). Accounting for higher
N(\ion{H}{1}), modest variations in ionization conditions, and
radiative transfer effects (the Lyman limit system has $\tau_{LL}
= N_{HI}/[1.59 \times 10^{17} {\rm cm}^{-2}] = 0.2$ at 1 ryd), it
is possible that these complex, multiphase absorbers have a
similar physical origin. For the low ions in the partial LLS,
Prochaska et al. (2004) derive $\log U = -2.9 \pm 0.2$ and $\log
Z/Z_{\odot} \gtrsim -0.3$, a metallicity similar to the range we
infer for the low ions in the PG~1211+143 15,300 \kms\ system.
Their model for this cloud is a multi-phase partial LLS composed
of photoionized gas in close association with a
collisionally-ionized \ion{N}{5} and \ion{O}{6} ``skin'' component
offset from the main \ion{H}{1} component.

In terms of their ionization states, column densities, linewidths,
and component structures, the PG~1211+143 15,300 \kms\ absorber and
PKS~0405-1219 LLS are strikingly similar. They resemble one another
in another critical aspect: both have bright galaxies at
$\rho_{\perp} \sim 100h_{70}^{-1}$ kpc. PKS~0405-1219 has an
elliptical galaxy at 75$h_{70}^{-1}$ kpc with $|\Delta v| = 120$
\kms\ and a spiral at 63$h_{70}^{-1}$ kpc with $|\Delta v| = 30$
\kms. Although incomplete galaxy redshift information in the
PKS~0405-1219 field hinders detailed conclusions, Chen \& Prochaska
(2000) conclude that both ``Galactic halo'' and ``intragroup
medium'' models are viable locations for CIE and PIE gas to produce
the observed column densities.  This ambiguity also arises in the
PG~1211+143 sightline. Improved galaxy/absorber statistics on
sightlines passing near galaxies, within and outside groups, are
necessary to determine the true physical origins of these intriguing
absorbers.

\subsection{Comparisons to 19,300 \kms}

The 19,300 \kms\ system toward PG~1211+143 shows strong \ion{C}{3},
weak \ion{Si}{3}, and two \ion{O}{6} components. While we cannot
conclusively rule out photoionization models for the 19,434 \kms\
\ion{O}{6} component if it is associated with the C component of
\ion{H}{1}, we favor an interpretation in which both \ion{O}{6}
components arise in CIE hot gas at $\log T \gtrsim 5.45$, with the
corresponding \ion{H}{1} blended with the stronger photoionized
systems. Regardless of the final interpretation, this system bears
strong resemblance to the \ion{O}{6} systems at $z = 0.1212$ toward
H1821+643 (Tripp et al. 2001) and at $z = 0.00337$ toward 3C~273
(Sembach et al. 2001; Tripp et al. 2002).  In all cases, broad
\ion{O}{6} is seen, with some evidence of broad associated
\ion{H}{1} and an additional, narrower \ion{H}{1} component that is
not aligned with the \ion{O}{6}.  The absence of metal-line
absorption other than \ion{O}{6} in the H1821+643 and 3C~273
absorbers (down to the detection limits of their respective studies)
may be a column-density effect; both systems have roughly one dex
lower N(\ion{H}{1}) than the PG~1211+143 19,300 \kms\ system. If
\ion{C}{3} and \ion{Si}{3} were present at the same metallicity and
ionization parameter as in the other clouds, they would have gone
undetected.  The 3C~273 absorber is located 169$h_{70}^{-1}$ kpc
from a galaxy at $|\Delta v| = 90$ \kms, and the H1821+643 system is
located 140$h_{70}^{-1}$ kpc from a galaxy at $|\Delta v| \lesssim
30$ \kms.

The PG~1211+143 19,300 \kms\ system also bears superficial
resemblance to the $z = 0.14232$ system toward PG~0953+415 (Savage
et al. 2002).  This system shows \ion{H}{1}, \ion{O}{6}, and (weak)
\ion{C}{3}. This system also lies in an apparent concentration of
galaxies, with the closest being a bright galaxy at $|\Delta v| \sim
130 $ \kms\ and offset $\sim$ 400$h_{70}^{-1}$ kpc from the
sightline (the Savage et al. 2002 galaxy survey did not probe below
$L \sim 0.5 L^*$). However, this absorber shows \ion{O}{6} with a
linewidth ($b = 13$ \kms) and column density ($\log N$(\ion{H}{1}) =
13.59) that can be explained by a reasonable photoionization model
with $\log (Z/Z_{\odot}) = -0.4$ and a 420 kpc line-of-sight
pathlength.  This system provides another interesting data point for
correlations of \ion{O}{6} and metals with galaxies, but it does not
parallel the PG~1211+143 systems as closely as do the others we
compare to here.

The most striking feature common to all these absorbers is the
velocity offset between the broad \ion{O}{6} components and the
nearby strong, narrow \ion{H}{1}.  In 3C~273 the broad \ion{O}{6}
lies in the wings of the photoionized \ion{H}{1} profile. In
PG~1211+143 the 19,356 \kms\ \ion{O}{6} is not clearly aligned with
any detected \ion{H}{1} component, and any associated \ion{H}{1}
should also be broader than the detected \ion{H}{1}. In H1821+143
and PG~1211+143, the velocity separations between the \ion{O}{6} and
one or more of the \ion{H}{1} components are consistent with an
origin in post- and pre-shock gas in a strong shock front between
the photoionized and \ion{O}{6}-bearing components, respectively.
The small velocity separation in the 3C~273 cloud could be
reconciled with this picture if the shock is viewed along its front
and the shock velocity vector lies near the plane of the sky. Also,
the absence of \ion{H}{1} Ly$\alpha$ absorption at $z \approx
0.00337$ in the spectrum of RXJ~1230.8+0115 (225$h_{70}^{-1}$ kpc
away) suggests a physical size to the \ion{O}{6} absorber smaller
than plausible PIE models, thus favoring CIE for the 3C~273 absorber
(Tripp et al. 2002).

Although there are some modest inconsistencies between these three
cases, they together suggest that \ion{O}{6} can arise in
collisionally ionized gas, shock-heated against lower ionization,
photoionized gas which may or may not show metal-line absorption.

\subsection{Extended Halos of Luminous Galaxies?}

To address the question of whether these multi-phase, metal-enriched
absorbers could be associated with the nearby galaxies, we review
the evidence for causal associations between galaxies and Ly$\alpha$
forest clouds.  In studies of Ly$\alpha$ absorption around galaxies,
Chen et al. (2001) and Bowen, Pettini, \& Blades (2002) concluded
that tenuous [$\log N$(\ion{H}{1}) $ \gtrsim 13$] gas exists around
galaxies out to $\rho_{\perp} \simeq 200h_{70}^{-1}$ kpc with a
covering fraction near 100\%. The PG~1211+143 systems lie within the
scatter in the relations of Ly$\alpha$ equivalent width and/or
$N$(\ion{H}{1}) vs. impact parameter presented by these two studies.
This consistency suggests that the PG~1211+143 systems may share a
common physical origin with the Ly$\alpha$ absorption studied by
these authors.

However, these previous studies differ in their physical
interpretation. Chen et al. (2001) concluded that the strong
correlation between $N$(\ion{H}{1}) and $\rho_{\perp}$ manifests a
causal connection between the absorption and individual galaxies.
They did not, however, evaluate the hypothesis that the \ion{H}{1}
traces not the individual nearby galaxies but rather the filamentary
large-scale structure in which the galaxies are also embedded. Bowen
et al. (2002) considered such a model and concluded that this latter
hypothesis is just as likely as association with individual
galaxies.  They showed that the correlation between Ly$\alpha$
equivalent width and galaxy density at the redshift of their
absorbers holds even if there is not a bright galaxy within
200$h_{70}^{-1}$ kpc of the sightline. Similar statistical
distributions of gas near galaxies have been reproduced in
hydrodynamical simulations that minimize the active ``feedback''
from galaxies that would be expected to cause such a strong
correlation (Dav\'{e} et al. 1999). Similarly, using a larger (46
absorber) sample discovered with {\em HST} in regions surveyed for
galaxies down to at least $L^*$, Penton, Stocke, \& Shull (2002)
found little statistical evidence that absorbers are associated with
individual galaxies. With a few notable exceptions (e.g., the dwarf
galaxy/absorber pair examined by Stocke et al. 2004a), a new study
by Stocke et al. (2005, in preparation) obtains the same result,
using a much larger absorber sample and extending the galaxy survey
down to $0.1L^*$ luminosity. Similar results were obtained by Morris
et al. (1993), Tripp, Lu \& Savage (1998), and Impey, Petry \& Flint
(1999). These findings suggest that most low column-density
absorbers are associated with large-scale galaxy filaments and not
with individual galaxies.  Thus, the current statistical evidence
is, at best, ambiguous on the point of whether Ly$\alpha$ clouds, or
a subset of the general Ly$\alpha$ forest population, are directly
associated with nearby galaxies. However, the arguments for or
against the association of Ly$\alpha$ clouds with galaxies cannot be
automatically extended to the subset that are also \ion{O}{6}
absorbers.

Unfortunately, the two absorbers toward PG~1211+143 only reinforce
this ambiguity. The sightline clearly grazes the edge of a
spiral-dominated group at 15,300 \kms\ and passes within
150$h_{70}^{-1}$ kpc of an apparently isolated galaxy at 19,300
\kms. These systems do, however, add one potentially critical piece
of information to the puzzle -- at least one component in each
system shows near-solar metallicity. This metallicity is seldom if
ever detected in the diffuse IGM, where $Z \sim 1 - 10$ \% solar is
more typical. This factor suggests that these absorbers lie within
the influence of the nearby galaxies, their smaller satellites, or
the tidal streams created by their interactions. We explore this
hypothesis we explore in \S~4.5.

Even if the observed \ion{H}{1} is not associated with the extended
halos of the nearby galaxies, the \ion{O}{6} might trace a hot, extended
coronal halo with $T \gtrsim 10^{6}$ K, as suggested for the Milky Way
by a variety of studies (see Sembach 2003 for a review).  For a hot,
collisionally ionized corona of radius $r$, a sightline passing through
its center will intercept a column density of \ion{O}{6} given by:
\begin{eqnarray}
N({\rm O~VI}) &  = & 2 (O/H)_{\odot}\; f_{OVI}\;  r\; Z\; n_H \nonumber \\
  & \leq & (2 \times 10^{12} \,{\rm cm}^{-2}) r_{150}
  Z_{0.1} n_{-5},
\end{eqnarray}
where we have scaled to halo radius $r_{150} = r/150$ kpc, density
$n_{-5} = n_H/10^{-5}$ cm$^{-3}$, and 0.1 solar metallicity, and
adopted $f_{OVI} \leq 3.8 \times 10^{-3}$ at halo temperatures $T
\geq 10^{6}$~K. This simple model is therefore unable to reproduce
the observed column density of $\log N$(\ion{O}{6}) $= 14.26$ at
15,341 \kms\ at the temperature and density expected for a hot
coronal galactic halo, even if the metallicity is solar.  We also
note that this simple model predicts $\log N$(\ion{H}{1}) $\leq
12.3$, so this hot halo would not appear in either the Ly$\alpha$
halo studies quoted above or in our data. As \ion{O}{7} and
\ion{O}{8} are more abundant in CIE at these temperatures, such a
scenario could be tested by X-ray observations of PG~1211+143 if
sensitivity to $N$(\ion{O}{7}) $\simeq 10^{14}$ cm$^{-2}$ could be
achieved.

Although the observed \ion{O}{6} does not provide direct evidence of
a hot, galactic halo at $\sim 150h_{70}^{-1}$ kpc from these
luminous galaxies, it may trace the interface between such a halo
and cooler infalling material, as has been proposed for a subset of
the local highly-ionized HVCs by Sembach et al. (2003) and Collins,
Shull, \& Giroux (2004). In this scenario, the \ion{O}{6} would
serve as indirect evidence of the hot coronal halo. We turn to this
issue in \S~\ref{hvcsection}.

\subsection{Extragalactic Analogs to High-Velocity Clouds?}
\label{hvcsection}

Since both PG~1211+143 cloud complexes contain \ion{O}{6} as well as other
low- and high-ionization lines, it is useful to compare column density
ratios found here with those found in \ion{O}{6}-bearing Galactic HVCs
(Sembach et al. 2003) for which additional ions are available. These
include the HVCs toward PKS 2155-304 and Mrk 509 (Collins et al. 2004),
and PG 1259+593 (Fox et al. 2004). We list their high-ion column density
ratios in Table~\ref{hvctable}, along with the measured values from the
two PG~1211+143 systems.

For consistency with the HVC analyses, we depart from our method of
fitting individual lines and instead integrate the apparent column
densities of the metal-line absorption over velocity ranges set by
the \ion{H}{1} absorption.  This apparent optical depth (AOD) method
is valid for unsaturated lines (Savage \& Sembach 1991) and
appropriate for most of the fitted metal lines, so the
high-ionization metal column-density ratios are not sensitive to
this alternative method\footnote{This analysis gives
model-independent measures of absorption in the overall line
profiles which are generally consistent with the column densities
derived from profile fitting in \S~3.}. These column densities
appear in Tables~\ref{15table} and \ref{19table}. While only limits
exist on the high-ion ratios for the 19,300 km s$^{-1}$ systems,
these limits suggest that this system is at least as highly ionized
as the one at 15,300 \kms.

This comparison shows that the ratios in the 15,300 \kms\ system
would not be out of place in the dataset including the other
\ion{O}{6} HVCs. The analyses of \ion{O}{6} HVCs by Collins et al.
(2004) and Fox et al. (2004) favor models in which the \ion{O}{6}
gas arises in conductive interfaces or turbulent mixing layers
between photoionized gas and hotter gas (cf. Table 7 in Fox et al.
2004).  The high-ion ratios in the 15,300 \kms\ system lie near the
region of parameter space occupied by shock ionization models and
conductive interface models compiled by Fox et al. (2004) and
Indebetouw \& Shull (2004a), and assuming solar abundance ratios.
The presence of lower ionization lines associated with the cooler
gas makes these identifications plausible, as the cooler gas would
be associated either with the pre-shock gas or the cooler side of
the conductive interface. Fox et al. (2004) argue that the high ions
associated with HVC Complex C arise in interfaces between the
neutral HVC and a hot ($\sim$ 10$^6$ K) Galactic halo. Models that
associate the \ion{O}{6} gas with turbulent mixing layers predict a
higher \ion{C}{4}/\ion{O}{6} ratio than the upper limits obtained
here (assuming solar C/O). Considering the ratios alone, they seem
more consistent with radiatively cooling gas flows, conductive
interfaces, or shock ionization (as argued in \S~3.2), indicative of
more dynamically active environments such as dwarf galaxy outflows
(see \S~\ref{outflowsection}).

However, an unequivocal explanation for the properties of the local
\ion{O}{6} HVCs remains elusive. Based upon the present data,
however, if such an explanation existed it could well be extended to
explain the gas in the PG~1211+143 absorbers. A common explanation
would rely on parallels between, for example, the PG~1211+143
systems and the low-$N$(\ion{H}{1}) ionized skin of Complex C or the
subset of Galactic \ion{O}{6} HVCs that do not show 21 cm emission
(Sembach et al. 2003). These systems illustrate that, first,
whatever mechanisms produce the Galactic highly-ionized HVCs, they
can operate more than 100 kpc from a massive galaxy, and second,
further searches near external galaxies should be undertaken to try
to understand HVCs from the perspective of a projected, rather than
radial, viewing geometry (e.g., Keeney et al. 2005). As the
PG~1211+143 absorbers would appear as HVCs from the perspective of
the nearby galaxies, they and systems like them could provide
important clues to the origins of the local highly ionized HVCs.

\subsection{Outflows from Nearby Galaxies?}
\label{outflowsection}

The two bright galaxies known to be near the PG~1211+143 sightline
suggest a model in which one or more of the absorption-line
components arises in material flowing out of the nearby galaxies,
perhaps in a starburst-driven superwind.  Stocke et al. (2004b)
review the literature on the potential connections between galaxies
and Ly$\alpha$ absorbers. They outline a general picture in which
most, and perhaps all, weak metal-line systems down to
$N$(\ion{H}{1}) $\simeq 10^{13-14}$ cm$^{-2}$ could arise in unbound
winds from dwarf galaxies like the one they found near 3C~273. In
this picture galaxies with $L \gtrsim 0.1 L^*$ are too massive for
starburst-driven winds to achieve escape velocity. This section
considers the PG~1211+143 absorbers in the context of this picture.

Two of the best examples of nearby, weak metal-line systems are
found in the sightlines to 3C~273 and RXJ~1230.8+0115, which are
separated by 0.91$^\circ$, or 363$h^{-1}_{70}$ kpc at $cz$ = 1600
\kms. Both sightlines have strong Ly$\alpha$ absorbers with
$N$(\ion{H}{1}) $ \approx 10^{16}$ cm$^{-2}$ at similar redshifts
($cz$ = 1586 and 1666 \kms, respectively) and metallicities (6\% and
10\% solar; Tripp et~al. 2002; Rosenberg et~al. 2003). The 3C~273
absorber best illustrates the galactic superwind model. Stocke
et~al. (2004a) discovered a dwarf ($M_B = -13.9$, or $L = 0.004
L^*$), post-starburst galaxy 70$h^{-1}_{70}$ kpc away from the 1586
km s$^{-1}$ absorber. This galaxy/absorber connection is among the
strongest in the literature:  the absorber and galaxy redshifts
match to within their combined errors, the absorbers and galaxy
metallicities match to within their larger errors, and the absorber
has an overabundance of silicon to carbon indicative of recent Type
II supernova enrichment.  The galaxy is a pure disk system showing
both strong Balmer and metal absorption lines in its optical
spectrum, and no evidence for dust or gas (no emission lines and
M$_{H I}\leq 5 \times 10^6$ M$_{\odot}$; van Gorkom et~al.  1993).
From ratios of Lick absorption-line indices, Stocke et al.  (2004a)
estimate that the mean stellar age in this galaxy is $3.5 \pm 1.5$
Gyr.

Taken together, this information provides a consistent picture in
which a massive ($\geq$ 0.3 $M_{\odot}$ yr$^{-1}$) starburst $\sim$
3 Gyr ago created enough supernovae to blow the galaxy's remaining
gas into the IGM. Because the dwarf is quite low in mass, this wind
can easily escape from the galaxy and move to $\sim 100h^{-1}_{70}$
kpc at the 20 - 30 \kms\ required to create the metal-line absorber
seen toward 3C~273.  This dwarf is estimated to have been $\sim$ 10
times brighter when starbursting. However, no dwarf galaxy with
similar properties or at comparable impact parameter has been found
near the RXJ~1230.8+0115 absorber at 1666 \kms. Thus, even though
the 3C~273 absorber establishes some of the necessary conditions for
a successful dwarf superwind model, there is not yet sufficient
evidence that this model is viable for all weak metal-line
absorbers, although a galaxy survey near RXJ~1230.8+0115 at a depth
comparable to the survey near 3C~273 has not been completed.

The dwarf galaxy superwind model does solve a difficult
geometrical problem emphasized by Rigby et al. (2002): the small
(10 pc - 10 kpc) sizes inferred for weak metal-line (\ion{Mg}{2},
\ion{Fe}{2}) absorbers are seemingly inconsistent with their high
frequency in QSO spectra ($dN/dz\approx$ 1-2 per unit redshift). A
local example of this quandary is the three metal-line absorbers
at comparable redshift in the 3C~273 and RXJ~1230.8+0115
sightlines. In this case, a single gaseous filament is probed by
two random sightlines $\sim 363h^{-1}_{70}$ kpc apart, requiring
high covering factor for metal-enriched ($Z \sim$ 2 - 10\% Solar)
gas along this filament. This is easily explained by the dwarf
superwind model because, while an individual superwind shell can
be quite thin along any sightline, the spherical expanding shell
subtends $\sim$ 100$h^{-1}_{70}$ kpc on the sky. Thus, only 3-5
dwarf superwind outflows would be required to completely cover the
area of this filament in metal-enriched gas. Because these dwarfs
are typically fainter than 0.1$L^*$, there would be little
statistical evidence for absorber-galaxy associations in $L^*$
galaxy samples (Bowen, Pettini, \& Blades 2002) but rather strong
evidence for absorber-galaxy filament associations, as has been
found (Penton, Stocke \& Shull 2002). This model also naturally
reconciles the very small line-of-sight sizes inferred from PIE
models for the metal-line absorbers in the 3C~273,
RXJ~1230.8+0115, and PG~1211+143 sightlines with the large
covering factors required to make this type of absorber relatively
common (see Stocke et al. 2004a; Rosenberg et al. 2003).

On the other hand, there is some evidence that more massive galaxies
($L \gtrsim 0.1 L^*$) retain their metal-enriched gas in
gravitationally-bound, optically-thick gaseous halos. Steidel (1995;
1998) found that almost all (53 of 55) strong \ion{Mg}{2} absorbers,
which are invariably optically-thick at the Lyman limit, have
nearby, bright ($L\gtrsim 0.1L^*$) galaxies within $\sim
50h^{-1}_{70}$ kpc on the sky. In addition, in a recent study of the
absorption lines associated with a luminous star-forming galaxy
($0.5L^*$, $SFR \simeq 1.4\, M_{\odot}$ yr$^{-1}$) galaxy NGC~3067,
Keeney et~al. (2005) show that the strongest absorber
[$N$(\ion{H}{1}) $= 10^{20.0}$ cm$^{-2}$] in the spectrum of the
quasar 3C~232 is not only gravitationally bound to the galaxy but is
most likely infalling. Taking these various results into account, we
would expect that the more luminous galaxies in the region of any
QSO sightline would retain their gas in $\sim 50h^{-1}_{70}$ kpc
bound halos.

When we consider the PG~1211+143 absorbers in the context of this
picture, the conclusions are not as clear. First, of the three
critical factors that enter the Stocke et al. (2004a) analysis,
only the proximity (in impact parameter and velocity) and
metallicity are available as discriminants for the PG~1211+143
absorbers. At least one of the components in each system has $Z
\simeq 0.5 - 1.0 Z_{\odot}$, and both lie within 150$h_{70}^{-1}$
kpc of bright galaxies.  Lick indices measured from
medium-resolution Las Campanas spectra (McLin 2002) indicate that
neither galaxy has undergone a starburst in the last 3 Gyr. Both
galaxies show old (3 - 8 Gyr), near solar-metallicity stellar
populations, with modest ongoing star formation (0.5 - 1.0
$M_{\odot}$ yr$^{-1}$) in the galaxy at 15,242 \kms.  This low
level of star formation in a galaxy of $L \sim L^*$ is unlikely to
have driven an unbound wind to $\sim 150h_{70}^{-1}$ kpc. We also
have no evidence for elevated Si/C ratios that would indicate
recent Type II SN enrichment\footnote{We cannot derive an accurate
estimate of the Si/C ratio at 15,300 \kms\ because photoionization
models that produce roughly one-to-one column-density ratios
$N$(C~III)/$N$(C~IV) and $N$(Si~III)/$N$(Si~IV) imply large and
highly uncertain ionization corrections for the unobserved ions
C~V, Si~V, and higher stages. Only a few lines of C and Si are
detected at 19,300 \kms, where the ionization correction is less
uncertain and [Si~III/C~III] $\simeq$ [Si/C] $\simeq -0.5$.}: the
Si/C ratio at 15,300 \kms\ is poorly constrained, and [Si/C]
$\simeq -0.5$ at 19,300 \kms . The two galaxies nearest to the
complex of PG~1211+143 absorbers are nearly $L^*$ galaxies ($0.7 -
0.8 L^*$ for $M^*_{B} = -19.6$ for $H_0 = 70$ km s$^{-1}$
Mpc$^{-1}$; Marzke et~al. 1994), and so would be expected to have
bound halos by this model. Since their impact parameters to
PG~1211+143 are twice as large as their expected halos, it is
unlikely that outflows from these galaxies are directly
responsible for the absorptions we see.

However, the 3C~273 case is relevant to the current cases if weak
metal-line absorbers arise in starburst winds from dwarf galaxies
fainter than our survey limit in this region ($m_B$ = 19 out to 150 or
$200h^{-1}_{70}$ kpc for these two absorbers, respectively). If placed at
15,000 - 19,000 \kms\ the dwarf near 3C~273 would have $m_B \sim 22-23$,
or $\sim 19.5 - 20.5$ while ``starbursting'', and it would therefore lie
below our survey limits. Limited attempts to observe fainter galaxies very
close to the PG~1211+143 sightline have not yet discovered dwarfs at the
absorber redshifts, but we cannot rule out their presence. Indeed, 20 -
30 dwarf galaxies with $L = 0.05 - 0.10 L^*$ would be expected at $\sim$
15,300 \kms\ given the number of brighter galaxies found in this galaxy
group (see Figure~\ref{pgmap}). If so, the photoionized components could
be due to one or more dwarf superwinds.  Colliding dwarf galaxy winds
from numerous nearby dwarf galaxies could be responsible for the O~VI,
since the typical shock velocity would reflect the velocity dispersion in
the group (100 - 200 \kms). Such a model would also naturally explain the
large number of components in the absorption-line systems. PG~1211+143
provides an opportunity to test this model for two absorbers with a
deeper galaxy survey in a single field.

\subsection{Intragroup medium at 15,300 \kms?}
\label{groupsection}

Even if the \ion{O}{6} at 15,242 \kms\ is not associated with the
nearby galaxy, it may still be associated with a hot intragroup
medium. As suggested by Mulchaey et al. (1996), spiral-rich groups
of galaxies are promising places to look for the predicted warm-hot
phase of the IGM. Spiral-rich groups are numerous and may be only
marginally bound. By scaling down from elliptical-rich groups with
soft X-ray detections, Mulchaey et al. (1996) suggested that if
spiral-rich groups possess intragroup gas, they should have physical
conditions just right to produce \ion{O}{6}. In this model,
spiral-dominated groups are too small and loosely bound to achieve
the high temperatures that would make them detectable in the soft
X-ray band. Thus, their predicted UV absorption lines would be the
best method of direct detection. Spiral-rich groups are numerous,
yet they occupy a small volume fraction of the universe. Thus, they
represent a potential reservoir for hot baryons that has not been
extensively probed by the undirected \ion{O}{6} surveys by {\em HST}
and {\em FUSE}.

The 15,300 \kms\ absorber toward PG~1211+143 is an excellent
candidate for the predicted hot intragroup medium.  The PG~1211+143
sightline grazes but does not pierce the spiral-rich group extending
$\sim 1$ Mpc away from the sightline to the northwest. Indeed, this
\ion{O}{6} absorber exhibits several features predicted by the
Mulchaey et al. (1996) model. First, the \ion{O}{6} has column
density $N$(\ion{O}{6}) $\simeq 10^{14}$ cm$^{-2}$, as predicted for
a typical group with $T \simeq 10^6$ K. The \ion{O}{6} is broad,
reflecting both thermal broadening and some contribution from bulk
(but perhaps unvirialized) motions in the group. The \ion{O}{6} is
too broad to be clearly associated with the \ion{C}{4} or
\ion{N}{5}, which would instead suggest an origin in photoionized
gas.  These factors suggest that the 15,300 \kms\ system does arise
in a hot intragroup medium. However, there are some significant
uncertainties in this picture. First, the \ion{O}{6} linewidth
($b_{OVI} = 53$ \kms) is narrower than the observed velocity
dispersion of the group. However, this may be expected, given that
the sightline grazes the edge of the group, rather than piercing the
middle.  A sightline that follows a chord through a roughly
spherical object should show less dispersion than a more central
sightline. The second uncertainty with an intragroup origin is that
the 19,300 \kms\ system shows two similar \ion{O}{6} components in
the absence of a spiral-rich group.  Thus, even if the intragroup
medium picture suggested by Mulchaey et al. (1996) is accurate, it
is clearly not a unique explanation for broad \ion{O}{6} absorption.
Nevertheless, we consider the intragroup medium model to be a viable
candidate for the origins of collisionally-ionized intergalactic
\ion{O}{6} absorbers. This model could be tested with better
statistics using sightlines that probe groups at different impact
parameters, and sightlines known to avoid groups.

If we adopt the hypothesis that the 15,300 \kms\ \ion{O}{6} arises
in a hot intragroup medium, we can ask whether this medium has any
relation to the photoionized gas in the other components at 15,300
\kms. In particular, do the lower-ionization components represent
partially recombined clumps or filaments that are pressure-confined
by the hotter low-density medium? At the peak ionization fraction of
\ion{O}{6} ($f_{OVI} = 0.22$ at $\log T = 5.45$), the gas pressure
$P/k = 2.24 n_HT$ can be calculated from the observed column density
of \ion{O}{6} and the line-of-sight extent $d$:
\begin{equation}
P/k = (1.9 \, {\rm cm}^{-3}\, {\rm K}) \left(\frac{N_{\rm
OVI}}{10^{14} \,{\rm cm}^{-2}}\right) \left(\frac{0.1Z_{\odot}}{Z}
\right) \left(\frac{1 \, {\rm Mpc}}{d}\right).
\end{equation}
We scale to one-tenth solar O abundance and a typical group size of
1 Mpc. For the 15,300 \kms\ \ion{O}{6} absorber, we derive $P/k
\simeq 3.8$ cm$^{-3}$ K for $\log N$(\ion{O}{6}) = 14.26 and
assuming $d = 900$ kpc and $Z = 0.1 Z_{\odot}$. The density implied
by this size is $n_H = 6 \times 10^{-6}$ cm$^{-3}$ at $\log T =
5.45$ K. Roughly one-half solar metallicity is required for this hot
gas to be in pressure equilibrium with the PIE component A (the only
moderate-ionization component that permits a detailed model), which
has 2\% metallicity. The factor of five discrepancy in pressure
between the two components could be erased by uncertainty in any of
the fiducial scaling parameters used in this calculation. However,
the component A cloud size of 180 -- 400 kpc would be difficult to
accommodate within a hot intragroup medium of Mpc scale. The large
photoionized component could still be a chance superposition near
but outside the group without negative implications for the group
model. The other photoionized components are also viable candidates
for pressure-confined low-ionization clouds, especially component C,
which shows $\gtrsim 40$\% solar metallicity. However, we lack the
necessary constraints on temperature and density to make detailed
models.  If the other PIE components have similar sizes, either a
chance location relative to this group or embedded clumps within the
group medium seem unlikely for all of them.

The hot intragroup medium model is viable (Equation 3) if $Z \simeq
Z_{\odot}$, $d \sim 1$ Mpc, $n_H = 2 \times 10^{-5}$ cm$^{-3}$, and
there is a $30 - 40$ \kms\ nonthermal contribution to the \ion{O}{6}
linewidth from bulk motions inside the group. The only factor in our
data that casts significant doubt on the intragroup medium model at
15,300 \kms\ is the presence of a similar multi-phase
Ly$\alpha$/\ion{O}{6} absorber at 19,300 \kms, where there is no
evidence of a spiral-dominated group. This qualitative discrepancy
is difficult to reconcile with the intragroup medium model for the
15,300 \kms\ system. In fact, the similarity between the 15,300
\kms\ and 19,300 \kms\ \ion{O}{6} absorbers makes it difficult to
argue that the nearby group exerts any influence at all on the
absorption complex at 15,300 \kms. Nevertheless, we find the hot
intragroup medium to be a viable model until additional evidence is
available (although it begs the question of the origin of the
metals).

A variant of the intragroup medium model was proposed by Shull, Tumlinson,
\& Giroux (2003) for the absorber complex at $cz \sim$ 17,000 \kms\ in the
PKS 2155-304 sightline. They proposed that the two \ion{O}{6} components
arise in warm photoionized gas shocked against the hot intragroup
medium filling the group potential. This model is consistent with our
findings for the physical conditions in the 15,300 \kms\ system, but it
is not unique in light of the 19,300 \kms\ system. If it is accurate,
it suggests that \ion{O}{6} may serve as a signpost of the true WHIM,
which could be verified by X-ray observations of \ion{O}{7}, \ion{O}{8},
and other species.

\subsection{Metal enrichment by continuous star formation?}

We have identified high-metallicity, photoionized material that appears
to have a recent connection to nearby galaxies. We also find \ion{O}{6}
components with velocity separations that suggest they trace the
post-shock regions of a strong shock against the photoionized gas.
However, connecting the \ion{O}{6} to the galaxies may extend the
chain of inference too far. As discussed in \S~4.4 and \S~4.6 There is
circumstantial evidence that these absorbers are at least associated
with the same large-scale structure that formed the nearby galaxies,
if not with the galaxies themselves. For a single sightline of $\Delta
z = 0.08$, two pairings of IGM absorbers with galaxies at $\rho_{\perp}
\lesssim 150h_{70}^{-1}$ kpc are extremely unlikely if they are random
(even more so in light of the PKS~0405-1219 case, see \S~4.2). These
absorbers also have stronger Ly$\alpha$ than expected from the
Penton et al. (2002) correlation of Ly$\alpha$ equivalent width with
$\rho_{\perp}$. Unfortunately we lack the necessary information (see
\S~4.6 and Stocke et al. 2004a) to establish firmer grounds for causal
relationships. However, if these two systems and the PKS~0405-1219 clouds
reside in filaments or intragroup medium and are not associated with
unseen satellites, their high level of metal enrichment is surprising.

At least one component in each of the PG~1211+143 absorber
complexes exhibits metallicity $Z \gtrsim 0.1 Z_{\odot}$, and the
highly-ionized \ion{O}{6} has 2 - 6 \% solar metallicity. Here we
estimate how much continuous star formation it takes to generate
$1 - 10$\% solar metallicity over the relevant range of physical
scales. We present two versions of the estimate. First, we scale
the estimate to the peak cosmic SFR density from all sources over
1 Gyr. Second, we scale to a Milky Way-like SFR ($\sim 1.0
M_{\odot}$ yr$^{-1}$) over a 150 kpc region. Both calculations
assume a mass yield in metals of $y_m = 0.024 \, M_{\odot}$ for
every $M_{\odot}$ formed into stars, and a baryon density
$\Omega_b h^2 = 0.0224$ (Spergel et al. 2003).

First, we estimate the total metal enrichment from all galaxies in
time $t = (10^9 {\rm yr})$ t$_9$, scaled to the peak cosmic SFR
density, $\rho(SFR)$. Then the IGM metallicity is given by:
\begin{eqnarray}
\frac{Z}{Z_{\odot}} & = & \frac{\rho(SFR) y_m t }{\Omega_b h^2 \rho_{cr} (0.02)} \nonumber \\
                    & = & \frac{(0.1 M_{\odot}\, {\rm yr}^{-1} {\rm Mpc}^{-3}) (0.024)
                                 (10^9\, {\rm yr}) t_9}{(0.0224)(1.879 \times 10^{-29} {\rm \,g\, cm}^{-3}) (0.02)} \nonumber \\
                    & = & (0.019) \left[ \frac{\rho(SFR)}{0.1 M_{\odot}\, {\rm yr}^{-1}\, {\rm Mpc}^{-3}}\right] t_9
\end{eqnarray}
Thus, $\sim 2$\% solar metallicity can be achieved in the IGM if star
formation proceeds for 1 Gyr at the peak rate of 0.1 $M_{\odot}$ yr$^{-1}$
Mpc$^{-3}$ measured at $z = 1 - 6$ (Steidel et al. 1999; Giavalisco et
al. 2004).

We can also estimate the metallicity within a volume of radius 150
kpc, such as that probed by the PG~1211+143 sightline in the halos
of the two nearby galaxies.  Within a 150 kpc radius, with mean
density $\left< n_H \right> = 10^{-4}$ cm$^{-3}$ and mean atomic
weight $1.32 m_H$, there is $4.6 \times 10^{10} M_{\odot}$ in gas.
A burst of star formation will generate a total mass in metals:

\begin{eqnarray}
M_{metals} & = & (1\, M_{\odot}\, {\rm yr}^{-1})\, y_m\, (10^9 \,{\rm yr})\, t_9 \nonumber \\
           & = & (2.4 \times 10^7 M_{\odot})\, t_9\, \left[ \frac{SFR}{M_{\odot}\, {\rm yr}^{-1}} \right]
\end{eqnarray}
Thus, the metallicity is:
\begin{eqnarray}
\frac{Z}{Z_{\odot}}&=&\frac{(2.4 \times 10^7 M_{\odot})\, t_9\, (SFR/M_{\odot}\, {\rm yr}^{-1})}{(4.6 \times 10^{10} M_{\odot})\, (0.02)} \nonumber \\
                   &=& (0.026) \left[ \frac{SFR}{M_{\odot}\, {\rm yr}^{-1}}\right] t_9
\end{eqnarray}
Thus $\sim$ 1 Gyr is required to achieve $2 - 3$\% solar metallicity with
steady star formation spread over large volumes.  This result suggests
that the lower metallicities of the photoionized components may be
achieved simply by the gradual buildup of metals from quiescent star
formation over 1 Gyr.  These absorbers are therefore not necessarily
associated with any particular galaxy; they may simply be embedded in
large-scale filaments of galaxies, as has been suggested for Ly$\alpha$
clouds and weak metal-line absorbers (Morris et al. 1993; Tripp, Lu \&
Savage 1998; Impey, Petry \& Flint 1999; Penton et al. 2002).  Such a
model does not naturally explain either the multi-phase nature of these
absorbers nor the small line-of-sight sizes for some of them.  Steady
metal enrichment seems insufficient to explain the higher metallicities
we see in component C (40\% solar) at 15,300 \kms\ and in A (40\% solar)
and D (90\% solar) at 19,300 \kms.  Rigby et al. (2002) and Charlton et
al. (2002) have emphasized this high metallicity problem in the context
of the ``weak-\ion{Mg}{2} absorbers'' found in ground-based spectra at
somewhat higher redshifts.  These high-metallicity absorbers seem to
require a single, metal-enriched unbound galaxy wind, metal-enriched
gas tidally stripped from an unseen satellite (see \S~4.4), or perhaps,
as proposed by Rigby et al. (2002), they are directly associated with
unseen stellar populations. Thus the gas metallicities inferred for
these multiphase absorption complexes suggest they are related to nearby
galaxies.

\section{DISCUSSION AND CONCLUSIONS} \label{conclusionsection}

We have used {\em HST}/STIS and {\em FUSE} to study two
multi-component, multiphase IGM absorption systems with substantial
\ion{O}{6}. These systems have the following important features, for
which any viable physical model must account:

\begin{itemize}
\item[1.] {\em They are multiphase:}
  The observed column densities of \ion{O}{6} in these systems are too
  high to be consistent with photoionization models based on the lower
  ions, except for a narrow range of parameter space that implies very
  large ($\sim$ Mpc) pathlengths together with substantial non-thermal
  broadening.  Although PIE models cannot be excluded conclusively, we
  believe that most of the \ion{O}{6} arises from collisional ionization
  in hot ($T = 100,000 - 300,000 $ K) gas. The detection of \ion{C}{3}
  - \ion{C}{4} and \ion{Si}{2} - \ion{Si}{4} together with \ion{O}{6} in
  these absorbers appears to trace multiphase structures composed of hot
  collisionally-ionized gas and warm photoionized gas possibly located
  close together or interacting. Dwarf galaxy outflows and, at 15,300
  \kms, intragroup medium models are both viable production mechanisms
  for the hot ionized gas.

\item[2.] {\em They are metal-enriched:}
   We find that all the metal-line systems here are enriched to at least
   $\sim 1$\% solar, and some have $Z/Z_{\odot} \gtrsim 40$\%. The small
   metallicity may reflect ``pre-enrichment'' by early generations of
   stars, but the higher metallicities suggest recent enrichment by
   galaxies by way of outflows or tidal stripping.

\item[3.] {\em They lie near galaxies and groups:}
   The discovery of these two multiphase absorbers in close association
   with galaxies and their resemblance to high-velocity clouds near the
   Milky Way (Collins et al. 2004; Fox et al. 2004) provides
   an important caution, or a critical clue, to the interpretation of
   \ion{O}{6} absorption in the general IGM. As pointed out by Sembach et
   al. (2003) for the {\em FUSE} survey of local \ion{O}{6}, it is unwise
   to interpret the ``IGM'' \ion{O}{6} as a single class of absorbers
   in the absence of detailed galaxy information and/or constraints on
   lower ionization species.

\item[4.] {\em There is no obvious ``unified model'':} The two
   strong absorption-line systems we study here are quite similar. Both show
   metal-enriched, multiphase gas, broad \ion{O}{6}, and both lie
   within $150h_{70}^{-1}$ kpc of $\sim L^*$ galaxies. However, they
   also show significant differences. One absorber is associated with
   a spiral-dominated group, while the other exhibits {\em two} strong
   \ion{O}{6} components.

\end{itemize}

While we have uncovered two absorbers potentially related to nearby
galaxies, there is no statistical evidence that this phenomenon is
common. The number of well-studied systems is far too small to allow
reliable conclusions about the general nature of \ion{O}{6}-bearing
hot gas near galaxies. Intensive searches in the $\lesssim$
200$h_{70}^{-1}$ kpc regions around galaxies and nearby QSO
absorbers should be undertaken to determine the extent to which our
results and those of Sembach et al. (2003) can be generally applied
to theories of galaxy formation and evolution. The dwarf-galaxy
outflow model can be tested by deeper galaxy surveys in this field,
but the intragroup medium model will require more high-quality data
on QSO sightlines passing through known galaxy groups.

As discussed by Tripp et al. (2001) for an \ion{O}{6} absorber of similar
strength in the sightline to H1821+643, the X-ray lines of \ion{O}{7}
and \ion{O}{8} might distinguish the true nature of the hot gas we
detect here, if present or future X-ray spectrographs can achieve
limiting column densities of $\log N$(\ion{O}{7}) $\simeq$ 14. Such a test
could eliminate the remaining PIE scenarios for the \ion{O}{6} absorbers
and better constrain their temperatures.  However, at the present time,
sensitivity to column densities as low as $\log N$(\ion{O}{7}) = 15
is available only in rare cases, such as when blazars flare (Mrk 421;
Nicastro et al. 2004). The study by Fang et al. (2002) using Chandra's
LETG show that this column density sensitivity will be difficult if not
impossible to achieve on a consistent basis in the near term, and may
require Constellation-X or later generation instruments.

\acknowledgements

We thank Brian Keeney for his help with the galaxy survey work and
Todd Tripp for comments that improved the manuscript.  This work is
based on data obtained for the Guaranteed Time Team by the
NASA-CNES-CSA {\em FUSE} mission operated by the Johns Hopkins
University. Financial support to U.S. {\em FUSE} participants has
been provided by NASA contract NAS5-32985. J. M. S. and M. L. G.
acknowledge the support of NASA LTSA grant NAG5-7262 and NSF grant
AST02-06042. J. T. S. and J. M. S. acknowledge support from HST
GO-8571.01-A, GO-8192.01-A, GO-6593.01-A, and AR-9221.01-A grants
for various studies of the low-$z$ Ly$\alpha$ forest.  J. T.
acknowledges support by the Department of Astronomy and Astrophysics
at the University of Chicago and HST GO-9874.01-A.  Based on
observations made with the NASA/ESA Hubble Space Telescope, obtained
at the Space Telescope Science Institute, which is operated by the
Association of Universities for Research in Astronomy, Inc., under
NASA contract NAS 5-26555.  Ground-based data were obtained with the
Apache Point Observatory 3.5-meter telescope, which is owned and
operated by the Astrophysical Research Consortium. We have made
extensive use of the NED, SIMBAD, and ADS online databases.

\clearpage

\clearpage

\clearpage

%\tabletypesize{\normalsize}
\begin{deluxetable}{cccccc}
\tablecolumns{6} \tablewidth{0pc} \tablecaption{Galaxies in the PG
1211+143 Field} \tablehead{RA & DEC & $m_B$\tablenotemark{a}  &
$cz$\tablenotemark{b}
           & $\rho_{\perp}$\tablenotemark{c} & Redshift \\
           (J2000) & (J2000)& (Source) &(\kms) &($h_{70}^{-1}$ kpc) & Source}
\startdata
12:14:19.9 & 14:05:10 & 17.9(3)   & 18,941  & 146  & Stocke \\
12:14:21.4 & 14:13:04 & 18.0(3)   & 14,888  & 575  & Stocke \\
12:13:39.5 & 14:14:40 & 17.7(2)  & 15,485  & 889  & CfA \\
12:13:51.8 & 14:13:23 & 15.7(1)   & 15,342  & 715  & CfA \\
12:14:00.6 & 14:11:39 & 17.5(2)  & 15,254  & 559  & CfA \\
12:14:01.1 & 14:11:08 & 18.0(2)  & 15,629  & 540  & CfA \\
12:14:06.3 & 14:12:55 & 16.8(2)  & 15,264  & 600  & CfA \\
12:14:07.4 & 14:09:24 & 15.3(1)   & 15,215  & 395  & CfA \\
12:14:09.5 & 14:04:23 & 17.3(3)   & 15,322  & 137  & CfA \\
12:14:17.7 & 14:03:14 & 14.6(1)   & 24,326  & 0    & PG 1211+143
\enddata
\tablenotetext{a}{B band apparent magnitude taken from (1) CfA, (2)
NED (http://nedwww.ipac.caltech.edu), or (3) Stocke et al. (2005, in preparation).}
\tablenotetext{b}{Heliocentric velocity.}
\tablenotetext{c}{Projected galaxy/sightline separations at the
     individual galaxy redshifts. These distances assume a flat universe
     with $H_0 = 70$ km s$^{-1}$ Mpc$^{-1}$,
    $\Omega_{M} = 0.3$, and $\Omega_{\Lambda} = 0.7$.}
\label{galtable}
\end{deluxetable}

\begin{deluxetable}{lccccc}
\tablecolumns{6} \tablewidth{0pc} \tablecaption{The 15,300 km
s$^{-1}$ System} \tablehead{ Ion/Line   &$cz$(km
s$^{-1}$)\tablenotemark{a} &$\log N$(cm$^{-2}$)  & $b$ (km s$^{-1}$)
& Group & $N(v)$\tablenotemark{b}} \startdata
\ion{H}{1} Ly$\alpha$ & 15,315$^{+8}_{-19}$& 15.21$^{+0.60}_{-0.50}$ & 36$^{+3}_{-7}$      &  A-C$^{*}$  & saturated    \\
                      & 15,407$^{+20}_{-7}$& 14.57$^{+0.05}_{-0.27}$ & 49$^{+4}_{-9}$      &  D          &  \nodata       \\
                      & 15,570$^{+3}_{-3}$ & 13.54$^{+0.05}_{-0.05}$ & 37$^{+3}_{-4}$      &  E          &  \nodata       \\
%                      &                   &                   &                           &             &                \\
\ion{H}{1} Ly$\beta$  & 15,323$^{+10}_{-10}$ & 15.60$^{+0.30}_{-0.10}$ & 32$^{+2}_{-5}$      &  A-C$^{*}$  & saturated      \\
                      & 15,435$^{+10}_{-10}$ & 14.20$^{+0.06}_{-0.06}$ & 45$^{+10}_{-10}$    &  D          & \nodata        \\
%                      &                   &                   &                           &             &                \\
\ion{H}{1} Ly$\gamma$ & 15,301$^{+10}_{-10}$ & 14.86$^{+0.10}_{-0.10}$  &22\tablenotemark{c} &  A          & 15.43$\pm$0.08 \\
                      & 15,349$^{+10}_{-10}$ & 15.31$^{+0.05}_{-0.05}$ & 21$^{+2}_{-2}$      &  B-C$^{*}$  & \nodata          \\
                      & 15,425$^{+10}_{-10}$ & 14.10$^{+0.65}_{-0.15}$ & 14$^{+10}_{-13}$    &  D          & \nodata         \\
%                      &                   &                   &                           &             &                   \\
\ion{C}{2} 1334       & 15,353$^{+3}_{-3}$ & 13.15$^{+0.10}_{-0.09}$ & 6$^{+2}_{-2}$       & C           & blend           \\
%                     &                   &                   &                           &             &                 \\
\ion{C}{3} 977        & 15,284$^{+10}_{-10}$ & 12.99$^{+0.05}_{-0.05}$ & 8\tablenotemark{d}  &  A          & 13.86$\pm$0.15 \\
                      & 15,337$^{+10}_{-10}$ & 13.92$^{+0.11}_{-0.04}$ & 20$^{+2}_{-3}$      &  B          &\nodata         \\
                      & 15,354$^{  }_{  }$\tablenotemark{e} & 14.14\tablenotemark{f} &  4$^{  }_{  }$    &  C/C$^{*}$  & \nodata \\
%                     &                   &                   &                           &             & \nodata         \\
\ion{C}{4} 1548       & 15,283$^{+3}_{-3}$ & 13.02$^{+0.08}_{-0.08}$ & 8$^{+4}_{-2}$       &  A          & 14.03$\pm$0.06 \\
                      & 15,319$^{+3}_{-3}$ & 13.08$^{+0.13}_{-0.12}$ & 9$^{+5}_{-2}$       &  \nodata    &\nodata          \\
                      & 15,340$^{+3}_{-3}$ & 13.75$^{+1.00}_{-0.17}$ & 6$^{+3}_{-4}$       &  B$^{*}$    &    \nodata     \\
                      & 15,354$^{+3}_{-3}$ & 13.43$^{+0.16}_{-0.27}$ & 5$^{+2}_{-3}$       &  C$^{*}$    &  \nodata        \\
%                      &                   &                   &                           &             &                 \\
\ion{Si}{2} 1260      & 15,354$^{+3}_{-3}$ & 12.25$^{+0.05}_{-0.05}$ & 5$^{+1}_{-1}$       &  C          & 12.29$\pm$0.12 \\
%                      &                   &                   &                           &             &  \nodata        \\
\ion{Si}{3} 1206      & 15,342$^{+3}_{-3}$ & 12.82$^{+0.05}_{-0.06}$ & 20$^{+2}_{-2}$      &  B          & 12.99$\pm$0.05 \\
                      %& 15,356$^{+3}_{-3}$ & 12.75$^{+1.53}_{-0.40}$ & 2$^{+2}_{-1}$      &  C          & \nodata        \\
                      & 15,356$^{+3}_{-3}$ & 12.37$^{+0.10}_{-0.10}$ & 5\tablenotemark{g}  &  C$^{*}$    &\nodata          \\
\ion{Si}{4} 1393      & 15,334$^{+3}_{-3}$ & 12.49$^{+0.08}_{-0.05}$ & 7$^{+1}_{-1}$       &  B$^{*}$    & 12.96$\pm$0.05  \\
                      & 15,353$^{+3}_{-3}$ & 12.67$^{+0.05}_{-0.07}$ & 5$^{+2}_{-1}$       &  C$^{*}$    & \nodata         \\
\ion{N}{5} 1242       & 15,341$^{+4}_{-5}$ & 12.98$^{+0.13}_{-0.20}$ & 15$^{+8}_{-8}$      &  B$^*$      &\nodata          \\
\ion{O}{6} 1038       & 15,341$^{+10}_{-10}$ & 14.26$^{+0.05}_{-0.08}$ & 53$^{+10}_{-9}$     & \nodata     & 14.21$\pm$0.08  \\
\enddata
\footnotesize
\tablenotetext{a}{Errors in velocity are constrained to $\geq$
one-half the resolution element;
                     $\pm 10$ \kms\ for {\em FUSE}, $\pm 3$ \kms\ for STIS/E140M.}
\tablenotetext{b}{Column density calculated from apparent optical
depth integration of the observed profiles over $v = $ 15,250-15,450 \kms.}
\tablenotetext{c}{This component was forced to have $b = 22$ \kms\
                   (\ion{H}{1} at $T \simeq 30,000$ K). If all parameters
                   are allowed to float, we obtain $\log N$(\ion{H}{1})
                   $= 14.77$, $b = 6$ \kms\ for component A, and $\log
                   N$(\ion{H}{1}) $= 15.38$ and $b  = 26$ \kms\ for
                   components B - C$^{*}$. The listed parameters are
                   adopted in the models in \S~3.}
\tablenotetext{d}{This \ion{C}{3} fit is poorly constrained. The tabulated
                   $N$ and $cz$ were fitted with a fixed $b = 8$ \kms\
                   from the \ion{C}{4} component A.}
\tablenotetext{e}{This velocity was fixed at \ion{Si}{2} component
C.} \tablenotetext{f}{The addition of this narrow \ion{C}{3} fixed
at 15,354 \kms\ does not change the
                  fits for A and B by more than their assigned errors,
                  but it carries uncertainty of $\sim
                  1.0$ dex in $N$(\ion{C}{3}). The listed parameters for
                  A and B do not include C; for C they are the overall
                  best fit with $cz$ fixed at 15,354 \kms.}
\tablenotetext{g}{This component fitted with $b = 5$ \kms\ fixed from
\ion{Si}{4} component C$^{*}$.} \label{15table}
\end{deluxetable}

\begin{deluxetable}{lccccc}
\tablecolumns{6} \tablewidth{0pc} \tablecaption{The 19,300 km
s$^{-1}$ System} \tablehead{ Ion/Line   &$cz$(km
s$^{-1}$)\tablenotemark{a} &$\log N$(cm$^{-2}$)  & $b$ (km
s$^{-1}$) & Group & $N(v)$\tablenotemark{b}} \startdata
\ion{H}{1} Ly$\alpha$ & 19,305$^{+3}_{-3}$ & 14.94$^{+0.07}_{-0.05}$& 34$^{+2}_{-2}$ & A+B       &   saturated \\
                      & 19,423$^{+3}_{-3}$ & 13.69$^{+0.05}_{-0.06}$& 27$^{+2}_{-2}$ & C         &   \nodata  \\
                      & 19,474\tablenotemark{c}& 13.40$^{+0.16}_{-0.22}$& 35$^{+11}_{-11}$ & D   &   \nodata   \\
                      & 19,560$^{+4}_{-5}$ & 12.87$^{+0.08}_{-0.10}$& 28$^{+6}_{-5}$ & E         &   \nodata \\
%                      &                   &                        &                &           &    \\
\ion{H}{1} Ly$\beta$  & 19,315$^{+10}_{-10}$ & 15.12$^{+0.06}_{-0.05}$& 30$^{+2}_{-3}$ & A+B       &   saturated \\
                      & 19,474\tablenotemark{d} & 13.65$^{+0.15}_{-0.30}$& 35\tablenotemark{d}& D&   \nodata \\
%                      &                   &                        &                &           &   \\
\ion{H}{1} Ly$\gamma$ & 19,278$^{+10}_{-10}$ & 14.46$^{+0.12}_{-0.05}$& 11$^{+2}_{-1}$\tablenotemark{e} & A & 15.18$\pm$0.06  \\
                      & 19,317$^{+10}_{-10}$ & 15.09$^{+0.06}_{-0.07}$& 21$^{+3}_{-3}$ & B         &    \nodata\\
%                      &                   &                        &                &           &   \\
\ion{C}{3} 977        & 19,277$^{+10}_{-10}$ & 13.89$^{+1.10}_{-0.70}$&  4$^{+5}_{-2}$\tablenotemark{f} & A         &   13.50$\pm$0.08\\
                      & 19,310$^{+10}_{-10}$ & 13.23$^{+0.10}_{-0.06}$& 27$^{+9}_{-6}$ & B         &   \nodata \\
                      & 19,474$^{+10}_{-10}$ & 12.96$^{+0.05}_{-0.04}$& 18$^{+6}_{-9}$ & D         &   13.02$\pm$0.08\\
%                      &                   &                        &                &           &   \\
\ion{C}{4} 1548       &  \nodata          & $\leq$ 12.5\tablenotemark{g}& \nodata   & \nodata    &   $< 13.1$\tablenotemark{h} \\
%                      &                   &                        &                &           &   \nodata\\
\ion{Si}{2} 1260      &  \nodata          & $\leq$ 11.5\tablenotemark{g}& \nodata   &  \nodata   &   \nodata\\
%                      &                   &                        &                &           &   \\
\ion{Si}{3} 1206      &  \nodata          & 12.20$^{+0.08}_{-0.08}$ & \nodata       & A          & \nodata   \\
%                      &                   &                        &                &           &   \\
\ion{Si}{4} 1393      &  \nodata          & $\leq$ 12.0\tablenotemark{g}&\nodata    & \nodata    &   $<12.6$\tablenotemark{h} \\
%                      &                   &                        &                &           &   \\
\ion{N}{5} 1242       & \nodata           & $\leq$ 12.2\tablenotemark{g}            & \nodata  &  \nodata     & $<12.8$\tablenotemark{h}  \\
%                      &                   &                        &                &          &    \\
\ion{O}{6}  1032      & 19,356$^{+10}_{-10}$ & 14.08$^{+0.05}_{-0.06}$& 37$^{+7}_{-6}$  &   \nodata         &  14.04$\pm$0.08 \\
                      & 19,434$^{+10}_{-10}$ & 13.63$^{+0.11}_{-0.20}$& 21$^{+10}_{-10}$     & C?      &
                      13.75$\pm$0.05 \\
\enddata
\footnotesize \tablenotetext{a}{Errors in velocity are constrained
to $\geq$ one-half the resolution element; $\pm 10$ \kms\ for {\em
FUSE}, $\pm 3$ \kms\ for STIS/E140M.} \tablenotetext{b}{Column
density calculated from apparent optical depth integration of the
observed profiles over $v =$ 19,300-19,490 \kms.  For O~VI the
integrals range over $v =$ 19,300-19,390 \kms\ and $v =$
19,390-19,490 \kms. For limits on \ion{C}{4}, \ion{Si}{4}, and
\ion{N}{5}, see note $h$.} \tablenotetext{c}{Velocity poorly
constrained; fixed at position of \ion{C}{3} D component.}
\tablenotetext{d}{For this component, the $N$ was fitted with $cz$
fixed at the velocity of \ion{C}{3} D component and $b$ fixed at the
value from the Ly$\alpha$ D component.} \tablenotetext{e}{The
b-value for this component was limited to $b > 10$ \kms\ to
accommodate the presence of \ion{C}{3} in photoionized gas at $T >
8000$ K.} \tablenotetext{f}{This formal best-fit linewidth is
strictly too narrow for FUSE to resolve. The formal error bar on
$N$(C~III) is large enough to overlap the value provided by direct
integration. Implications for the absorber model are discussed in
\S~3.2.} \tablenotetext{g}{4$\sigma$ upper limits, signal-to-noise
determined in 2-pixel resolution elements ($R = 44,000$).}
\tablenotetext{h}{4$\sigma$ upper limits, integrated over 100 \kms\
velocity range for comparison to detected O~VI lines in CIE. See
Section 3.2 for details.} \label{19table}
\end{deluxetable}

\begin{deluxetable}{ccc}
\tablecolumns{3} \tablewidth{0pc} \tablecaption{Metallicities for
the Detected Components}
 \tablehead{Label & $v$ (km s$^{-1}$)& $Z/Z_{\odot}$}
\startdata \cutinhead{15,300 \kms}
A & 15,288  & 0.02 $\pm$ 0.005  \\
B-C$^*$ & 15,340-15,354   & $\gtrsim 0.40$\tablenotemark{a}     \\
\cutinhead{19,300 \kms}
A & 19,274  & $\geq 0.4 $  \\
B & 19,317  & $0.05 - 0.25$    \\
D & 19,474  & $\geq 0.6 $ \\
O~VI & 19,356   & $\geq 0.06$  \\
O~VI & 19,434  & $\geq 0.02$
\enddata
\tablenotetext{a}{Estimate for blended components B,B$^{*}$, C,
and C$^{*}$ derived from photoionization considerations. See
\S~3.1 for discussion.} \label{metaltable}
\end{deluxetable}

\begin{deluxetable}{cccccc}
\tablecolumns{6} \tablewidth{0pc} \tablecaption{Shock models for
19,300 \kms\ O~VI} \tablehead{Label & $\Delta v$\tablenotemark{a} &
$\log T_s$ & $b_{th}$ & $c_s$\tablenotemark{b} & Mach \\ & (\kms) &
(K) & (\kms) & (\kms) & number } \startdata \cutinhead{19,356 \kms}
A & +137  & 5.41 & 16 & $14-20$ & $7 - 10$  \\
B & +68   & 4.79 & \nodata   & \nodata      & \nodata        \\
C & $-116$  & 5.27 & \nodata   & \nodata      & \nodata        \\
D & $-204$  & 5.76 & 24 & $18-22$ & $9 - 11$  \\
E & $-353$  & 6.24 & 42 &  $\sim$33   &  $\sim$11 \\
\cutinhead{     19,434 \kms}
A & +272  & 6.00 & 32   & $14-20$ & $14 - 19$ \\
B & +202  & 5.75 & 24   & $14-30$ &  $7 - 14$ \\
C & +19   & 3.70 & \nodata &\nodata &\nodata \\
D & $-69$   & 4.82 & \nodata &\nodata &\nodata \\
E & +218  & 5.82 & 26   &  $\sim$33   &  $\sim$7\\
\enddata
\tablenotetext{a}{Scaled up by $\sqrt{3}$ to account for average 3D
to 1D projection effects.} \tablenotetext{b}{Adiabatic sound speed
$c_s \equiv (\gamma k T/\mu)^{1/2}$ with $\mu = 0.6m_H$ and $\gamma
= 5/3$.} \label{19shocktable} \end{deluxetable}

\begin{deluxetable}{lcccc}
\tablecolumns{5}
\tablewidth{0pc}
\tablecaption{Comparisons to Galactic O~VI HVCs}
\tablehead{System & $\log\left(\frac{{\rm Si~IV}}{{\rm O~VI}}\right)$ &
                    $\log\left(\frac{{\rm C~IV}}{{\rm O~VI}}\right)$  &
                    $\log\left(\frac{{\rm N~V}}{{\rm O~VI}}\right)$  & Note}
\startdata
PKS~2155-304 & $-1.02 \pm 0.08$ & $ -0.32 \pm 0.06$  & $<-0.72$  & 1  \\
PKS~2155-304 & $-1.30 \pm 0.16$ & $0 \pm 0.06$       & $<-0.47$  & 2  \\
Mrk 509     & $< -1.36       $ & $-0.36 \pm 0.06$   & $<-0.81$  & 3  \\
Mrk 509     & $-0.4 \pm 0.2  $ & $0.22 \pm 0.02$    & $<-0.69$  & 4  \\
PG~1259+593  & $-1.0 \pm 0.1    $ & $0.46 \pm 0.06$    & $<1.15$   & 5  \\
PG~1211+143  & $-1.25 \pm 0.07$ & $-0.18\pm0.07$      & $-1.23\pm0.12$   & 6 \\
PG~1211+143  & $<-1.4$          & $<-0.9$            & $<-1.2$   & 7  \\
PG~1211+143  & $<-1.2$          & $<-0.7$            & $<-1.0$   & 8  \\
\enddata
\label{hvctable} \tablecomments{(1) Component 1 ($V_{LSR} = -140$
\kms) toward PKS~2155-304 (2) Component 2 ($V_{LSR} = -270$ \kms)
toward PKS~2155-304 (3) Component 1 ($V_{LSR} = -240$ \kms) toward
Mrk 509 (4) Component 2 ($V_{LSR} = -300$ \kms) toward Mrk 509. The
PKS~2155-304 and Mrk 509 data are from Collins, Shull, \& Giroux
2004. (5) Fox et al. 2004 (6) integrated column density over
$v =$ 15,250 - 15,450 \kms\ (7) integrated column density over $v =$ 19,300 -
19,390 \kms\ (8) integrated column density over $v =$ 19,390 - 19,490
\kms.}
\end{deluxetable}

\end{document}